\documentclass[11pt]{article}

\usepackage[margin=1in]{geometry}
\usepackage{microtype}

\usepackage{authblk}

\usepackage{amsmath, amssymb, amsthm, mathtools}

\usepackage[justification=raggedright]{caption}
\usepackage{subcaption}
\usepackage{graphicx}
\usepackage{float}

\graphicspath{{./}{./figures/}}

\usepackage{booktabs}
\usepackage{siunitx}

\usepackage{hyperref}
\hypersetup{
  pdftitle={Projected Hessian Learning: Fast Curvature Supervision for Accurate Machine-Learning Interatomic Potentials},
  pdfauthor={Austin Rodriguez},
  colorlinks=true,
  linkcolor=blue,
  citecolor=blue,
  urlcolor=blue
}

\usepackage[numbers,sort&compress]{natbib}

\usepackage{enumitem}

\usepackage{xcolor}

\usepackage[ruled,vlined]{algorithm2e}

\title{\textbf{Projected Hessian Learning: Fast Curvature Supervision for Accurate Machine-Learning Interatomic Potentials}}
\author[1,2]{Austin Rodriguez}
\author[3]{Justin S.\ Smith}
\author[1]{Sakib Matin}
\author[4,*]{Nicholas Lubbers}
\author[1,5,*]{Kipton Barros}
\author[2,6,*]{Jose L.\ Mendoza-Cortes}

\affil[1]{Theoretical Division, Los Alamos National Laboratory, Los Alamos, New Mexico 87545, United States.}

\affil[2]{Department of Chemical Engineering \& Materials Science, Michigan State University, East Lansing, Michigan 48824, United States.}

\affil[3]{NVIDIA Corp., 2788 San Tomas Expy, Santa Clara, California 95051, United States.}

\affil[4]{Computer and Artificial Intelligence Division, Los Alamos National Laboratory, Los Alamos, New Mexico 87545, United States.}

\affil[5]{Center for Nonlinear Studies, Los Alamos National Laboratory, Los Alamos, New Mexico 87545, United States.}

\affil[6]{Department of Physics and Astronomy, Michigan State University, East Lansing, MI, 48823, United States.}

\affil[*]{Corresponding authors: \texttt{jmendoza@msu.edu}, \texttt{kbarros@lanl.gov}, \texttt{nlubbers@lanl.gov}}

\date{\today}

\begin{document}
\maketitle

\begin{abstract}
The Hessian matrix of the second derivatives contains substantially richer information about the local geometry of the potential energy surface than energies and forces alone. Although incorporating full Hessians into machine-learning interatomic potential (MLIP) training can significantly improve accuracy and robustness, the quadratic computational and memory cost of explicitly constructing and storing Hessian matrices has limited its practical use. 

Here, we introduce \textit{Projected Hessian Learning} (PHL), a scalable second-order training framework that incorporates curvature information using only Hessian–vector products (HVPs). By avoiding explicit Hessian construction and instead projecting curvature along stochastic probe directions, PHL reduces the cost of second-derivative supervision to near force-level complexity. The resulting stochastic loss, based on a Hutchinson trace estimator, is unbiased and exhibits favorable scaling with system size, enabling curvature-informed training without quadratic memory growth.

We evaluated various training approaches on a chemically diverse dataset of reactants, products, transition states, intrinsic reaction coordinates, and normal-mode sampled geometries generated at the theoretical level $\omega$B97XD / 6-31G(d). Four training schemes are considered: energy-force training (E-F), two HVP-based approaches (E-F-HVP with one-hot or Hutchinson vectors) and energy-force-Hessian training (E-F-H). When probe vectors are randomized for each minibatch, both HVP-based approaches achieve statistically indistinguishable energy, force, and Hessian accuracy relative to full Hessian training while providing more than $24\times$ speedups per epoch for the small molecular systems studied here. In a more realistic fixed-vector regime with only one HVP per molecule, Hutchinson projections consistently outperform one-column probing, particularly for far-from-equilibrium geometries.

Overall, PHL effectively replaces explicit Hessian supervision with force-complexity curvature training, retaining most of the accuracy benefits of full second-order methods while enabling scalable MLIP development for larger and more complex molecular systems.
\end{abstract}


\section{Introduction}
\begin{figure}[ht]
    \centering
        \includegraphics[width=\textwidth]{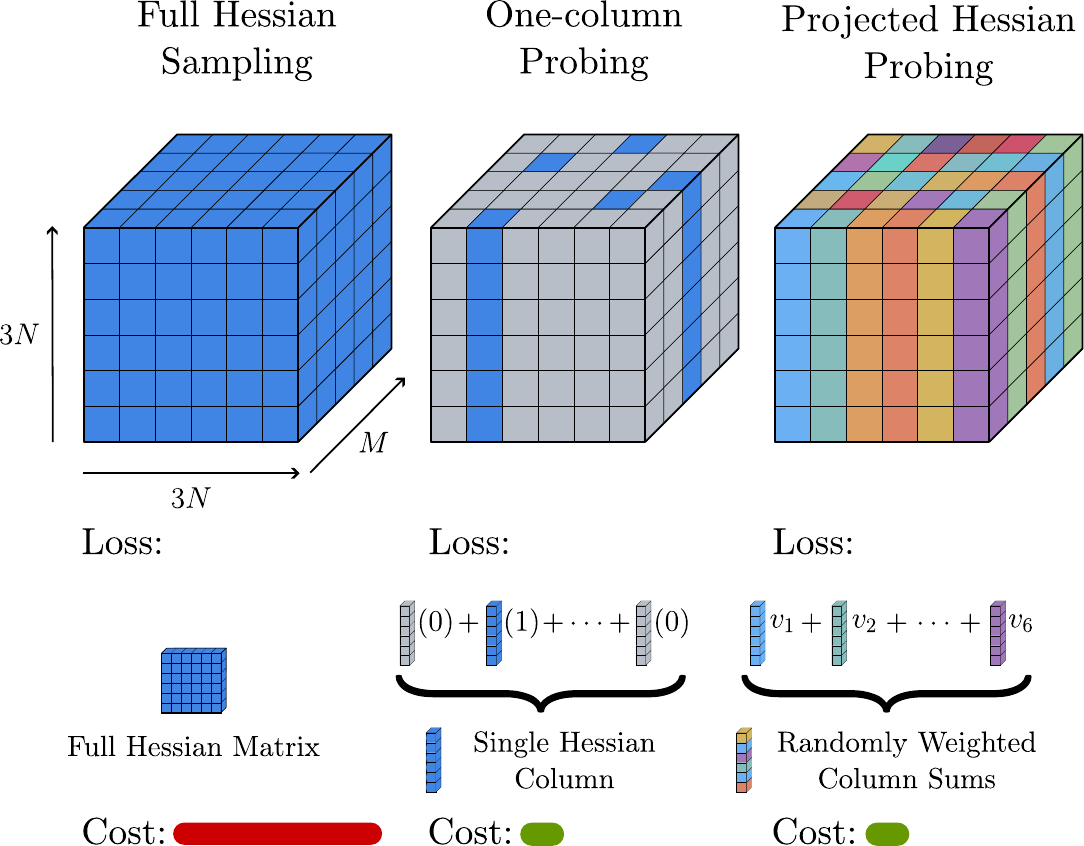}
    \caption{Conceptual comparison of curvature supervision strategies used to train MLIPs. Left: Full Hessian training explicitly uses all $(3N)\times(3N)$ second-derivative elements of $M$ systems. Middle: One-column HVP probing samples a single Hessian column via a canonical basis vector. Right: Projected Hessian Learning (PHL) (this work). Curvature is enforced through stochastic projections using random probing vectors to form Hessian-vector products, yielding random weighted combinations of Hessian columns (Hutchinson-style probing). Both HVP-based strategies avoid explicit Hessian construction; PHL aggregates information across multiple curvature directions in expectation.
    }
    \label{fig:main_figure}
\end{figure}

The quality of training data often imposes a fundamental limit on the accuracy of a machine learning interatomic potential (MLIP).\cite{unke2021machine} In neural-network–based MLIPs, reference data are typically obtained from quantum chemistry calculations such as density functional theory (DFT) or Coupled Cluster (CC), and the ML model is trained to reproduce these reference energies while enabling orders-of-magnitude faster inference.\cite{unke2021machine} A key insight established across multiple MLIP frameworks is that including gradients of the energy with respect to atomic coordinates (i.e. atomic forces) in the training loss substantially improves both accuracy and generalization. This paradigm has been realized in a wide range of architectures, including Behler–Parrinello neural networks,\cite{behler2007generalized} ANI models,\cite{Smith2017-1,Smith2017-2,Smith2018,Smith2020} HIP-NN,\cite{lubbers2018hierarchical} the DeepMD framework,\cite{zhang2018deep} kernel/descriptor-based MLIPs such as Gaussian Approximation Potentials (GAP),\cite{bartok2010gaussian} Spectral Neighbor Analysis Potentials (SNAP),\cite{thompson2015spectral} and Moment Tensor Potentials (MTP),\cite{shapeev2016moment} as well as message-passing approaches such as SchNet and NequIP.\cite{schutt2018schnet,batzner2022NequIP} More recent equivariant message-passing architectures, e.g.\ PaiNN and MACE, \cite{schutt2021equivariant,batatia2022mace} further reinforce the central role of force supervision in improving data efficiency and robustness. In all of these works, joint training on energies and forces consistently yields lower prediction errors and improved robustness compared to training on energy alone,\cite{chmiela2019sgdml,unke2019physnet} reflecting the strong regularizing effect and the enhanced efficiency of the data provided by the first-derivative information. While energy and force training has become standard, response properties that depend explicitly on second derivatives can remain inaccurate even when energies and forces are well reproduced. In crystalline materials, for example, harmonic phonons are determined by the (mass-weighted) force-constant matrix, i.e., the energy Hessian, and recent benchmarks show that several high-performing universal MLIPs still exhibit substantial errors in phonon observables despite strong energy and force accuracy.\cite{loew2025universal}

A well-implemented quantum chemistry code can produce not just the total energy but also the gradient of energy with respect to atom positions. Indeed, the principle of reverse-mode automatic differentiation,\cite{griewank2008evaluating,baydin2018automatic} known as backpropagation in machine learning, ensures that all force components can be calculated at a cost comparable to that of the energy evaluation alone; in electronic structure theory this is closely related to analytic energy-derivative techniques used to obtain nuclear forces.\cite{pulay2014analytical} Consequently, it is now standard for DFT codes to provide forces in addition to energies. The situation is different, however, for the Hessian matrix of second derivatives,
\begin{equation}
H_{(i,\alpha),(j,\beta)}=\frac{\partial^{2}E}{\partial r_{i}^{\alpha}\partial r_{j}^{\beta}},
\end{equation}
where $r_{i}^{\{x,y,z\}}$ denotes components of the position of the atom $i$. Elements of the Hessian can be understood as the sensitivity of the force on an atom $i$ with respect to perturbations to the position of an atom $j$. Calculating all such sensitivities is substantially slower than calculating forces alone, as analytic second derivatives generally require additional response equations beyond those used for gradients.\cite{pople1979derivative,pulay2014analytical}

For a system of $N$ atoms, the total energy is a single scalar, whereas the full gradient is a $3N$-dimensional vector, dramatically expanding the informational content available for training. This naturally raises the question of whether second derivatives of the energy, i.e.\ curvature information, can further improve the accuracy and generalization of the MLIP. The second derivative matrix, or Hessian, contains $(3N)^2$ components and governs the local geometry of the potential energy surface (PES), including vibrational frequencies, transition-state curvature, and reaction pathways.\cite{pulay2014analytical,gonzalez1989improved,fukui1981path} Motivated by these physical insights, recent work has begun to explore Hessian-informed approaches in ML potential development, including direct Hessian learning,\cite{rodriguez2025hessian} analytical ML Hessians for the characterization of transition-states,\cite{yuan2024analytical} Hessian distillation frameworks for specialized force fields,\cite{amin2025towards} and the use of Hessians as diagnostic tools to assess force locality in MLIPs.\cite{herbold2022hessian} Very recently, Koker \emph{et al.}\ proposed phonon fine-tuning (PFT), which directly supervises second-order force constants by matching MLIP energy Hessians to DFT-derived force constants from finite-displacement phonon workflows.\cite{koker2026pft} In a complementary direction, Burger \emph{et al.} introduced Hessian Interatomic Potentials (HIP), which predict symmetry-preserving Hessians directly from equivariant message-passing features without relying on automatic differentiation or finite differences.\cite{burger2025shoot} 

Complementing these methodological developments, the availability of \emph{curvature-labeled} datasets is rapidly expanding. For equilibrium molecular structures, \emph{Hessian QM9} provides numerical DFT Hessians for 41{,}645 QM9 molecules (including vacuum and implicit-solvent environments), enabling systematic evaluation of Hessian-aware training for vibrational properties.\cite{williams2025hessian} For reactive chemistry, the recently released \emph{HORM} database scales this idea dramatically, reporting 1.84 million quantum-chemistry Hessian matrices for diverse reactive geometries and transition-state workflows.\cite{cui2025large} The dataset used in this work complements these resources by targeting chemically diverse reaction pathways with equilibrium reactants, products, and transition states (RTP), continuous intrinsic reaction coordinate trajectories (IRC), and far-from-equilibrium normal-mode perturbations (NMS), providing a controlled arena to assess how curvature supervision impacts both interpolation and extrapolation in reactive MLIPs.\cite{rodriguez2025data}

Despite this promise, incorporating full Hessian information presents substantial practical challenges. First, acquiring Hessian data from quantum chemistry calculations can be prohibitively expensive for many electronic structure methods, particularly beyond mean-field levels where analytic second derivatives are substantially more complex or unavailable.\cite{pople1979derivative,pulay2014analytical} Second, training MLIPs with explicit Hessian supervision is significantly more expensive than energy–force training alone.\cite{rodriguez2025hessian} Although gradients can be computed efficiently using reverse-mode automatic differentiation, there is no general method to form all Hessian elements with comparable efficiency.\cite{griewank2008evaluating} Moreover, storing the Hessian requires $(3N)^2$ floating-point values per structure, leading to quadratic memory scaling with system size, which rapidly becomes a bottleneck for both data generation and GPU-based neural network training. These combined computational and memory costs severely limit the direct use of full Hessians in large-scale MLIP workflows, motivating alternative strategies that retain essential curvature information without explicitly constructing the Hessian.

In this work, we introduce \emph{Projected Hessian Learning} (PHL) as a stochastic strategy to efficiently incorporate curvature information into the training of machine learning interatomic potentials (Fig.~\ref{fig:main_figure}). The central idea in PHL is to avoid explicit construction of the full Hessian matrix and instead supervise \emph{projected} curvature information through Hessian-vector products (HVPs) of the learned potential energy surface, as schematized by the randomized probing strategy in Fig.~\ref{fig:main_figure}. The complete PHL training loop is summarized in Appendix~\ref{app:phl_algorithm}. Within an automatic differentiation framework used to train MLIPs, HVPs can be efficiently computed using forward-over-reverse or reverse-over-reverse differentiation without explicitly forming the Hessian.\cite{pearlmutter1994fast, griewank2008evaluating} HVP-based methods are also foundational in large-scale second-order optimization (e.g.\ Hessian-free and Newton conjugate gradient methods) precisely because they provide curvature access without forming Hessians.\cite{martens2010deep,nocedal2006numerical} In terms of floating-point operations, a single HVP can often be evaluated at a cost comparable to a small constant number of gradient evaluations, regardless of the size of the system $N$.\cite{pearlmutter1994fast,nocedal2006numerical} Finally, PHL’s use of low-dimensional probing subspaces is closely related to randomized “sketching” and low-rank approximation ideas for implicitly accessed matrices.\cite{halko2011finding,woodruff2014sketching}

\section{Mathematical Background}

\subsection{Stochastic estimation of the Hessian loss}

We can use the Hutchinson trace estimator to work with Hessian-vector products instead of the expensive calculation of the full Hessian matrix.\cite{hutchinson1990stochastic} This estimator says that given any matrix $A$, one can form an unbiased approximation to the trace using only the matrix-vector product;

\begin{equation}
\mathrm{tr}\,A\approx v^{T}Av.
\label{eq:stoch_tr}
\end{equation}

Here, $v$ is a random vector sampled from a distribution with components that are not correlated and have unit variance;

\begin{equation}
\left\langle v_{i}v_{j}\right\rangle =\delta_{ij}.
\label{eq:unitvar}
\end{equation}

The unbiased nature of this approximation can be verified by applying the definition of the Kronecker symbol $\delta_{ij}$ and the linearity of the expectation values $\left\langle \cdot\right\rangle $,

\begin{equation}
\mathrm{tr}\,A=\sum_{i}A_{ii}=\sum_{ij}\delta_{ij}A_{ij}\approx\sum_{ij}\left\langle v_{i}v_{j}\right\rangle A_{ij}=\left\langle v^{T}Av\right\rangle .\label{eq:unbiased}
\end{equation}

A simple and reasonable choice that satisfies Eq. (\ref{eq:unitvar}), as suggested by Hutchinson, is to independently randomize each component $v_{i}=\{\pm1\}$. Other choices are possible and will be considered below.

Equation \ref{eq:stoch_tr} can be incorporated into MLIP training as follows: First, consider a training loss function that involves energy, force, and Hessian loss terms,

\begin{equation}
\mathcal{L}=\lambda_{E}\mathcal{L}_{E}+\lambda_{F}\mathcal{L}_{F}+\lambda_{H}\mathcal{\mathcal{L}}_{H}.
\end{equation}

The hyperparameters $\lambda_{\{E,F,H\}}$ describe the relative weighting of these loss terms. Frequently, the $l2$ loss on components would be used. For example, a molecule of $N$ atoms may appear in the loss as

\begin{equation}
\mathcal{\mathcal{L}}_{H}=\frac{1}{(3N)^{2}}\sum_{i,j=1}^{3N}\left(\tilde{H}_{ij}-H_{ij}\right)^{2}.\label{eq:LH}
\end{equation}

The symbol $\tilde{H}_{ij}$ will denote the Hessian of the ML-predicted energy $\tilde{E}$, which has an implicit dependence on the model parameters $\theta$. The symbol $H_{ij}$ will denote true (reference) quantum chemistry Hessian data. It is our goal to train the model parameters to minimize the loss. In particular, stochastic gradient descent requires one to calculate $\nabla_{\theta}\mathcal{L}_{H}$. Clearly, this seems to require performing the complete sum on the components of the $(3N)^{2}$ matrix $\tilde{H}_{i,j}$.

Let us now review the construction of an efficient stochastic approximation. Equation (\ref{eq:LH}) can be interpreted as a matrix trace,

\begin{equation}
\mathcal{L}_{H}=\mathrm{tr}A.
\end{equation}

using the following definitions:
\begin{align}
A & =B^{T}B\\
B & =\frac{\tilde{H}-H}{3N}.\label{eq:B}
\end{align}

The matrix $B$ denotes the squared error in the Hessian predicted by the ML model $\tilde{H}$. One could also write $A=B^{2}$ since Hessian matrices are symmetric (i.e. derivatives with respect to atom positions commute). From the matrix trace, one can readily obtain a stochastic approximation,

\begin{equation}
\mathcal{L}_{H}\approx \hat{\mathcal{L}}_H = v^{T}Av=|Bv|^{2}.
\end{equation}

Back-substitution yields an explicit approximation formula,

\begin{equation}
\mathcal{L}_{H}\approx\frac{1}{(3N)^{2}}|\tilde{H}v-Hv|^{2},\label{eq:L_hutchinson}
\end{equation}
where $v$ is a random vector of component $3N$ that satisfies Eq. (\ref{eq:unitvar}). Importantly, the right-hand side involves only Hessian-vector products.\cite{pearlmutter1994fast,griewank2008evaluating} In particular, individual elements of the ML-predicted Hessian $\tilde{H}$ never need to be calculated explicitly. Instead, one can employ, e.g., the \texttt{torch.hvp} function to efficiently evaluate $\tilde{H}v$ as a vector.

\subsubsection{Randomizing over mini-batches}

Generally, a stochastic Hutchinson estimator such as Eq. \ref{eq:L_hutchinson} will be deployed as an average over a large number $N_{v}$ of random vectors $v$. Since each random vector is associated with an independent and unbiased estimate, the total stochastic error in estimating $\mathcal{L}_{H}$ decays like the inverse square root, $N_{v}^{-1/2}$.\cite{avron2011randomized,saibaba2017randomized}

In practice, training of MLIPs uses stochastic gradient descent (SGD) over a very large number of mini-batches. It is natural that each mini-batch could employ a new random vector $v$. Then, for purposes of SGD training, the overall quality of the Hutchinson approximator can benefit strongly from having many mini-batches.

\subsubsection{Random vector distributions}

As discussed above, Hutchinson’s original proposal was to sample each vector component independently under a uniform distribution,

\begin{equation}
v_{i}^{\textrm{Hutch}}=\{\pm1\}.
\label{eq:hutch}
\end{equation}

Recently, in the context of Hessian training, an alternative stochastic estimator was suggested: Work with a randomly selected column of $(\tilde{H}-H)$ at a time.\cite{koker2026pft,amin2025towards,cui2025large} This method can be cast into our notation taking $v$ to have a single non-zero component (“hot”) at a random index $c=\{1,\dots,3N\}$. This one-hot distribution can be expressed concisely as follows.

\begin{equation}
v_{i}^{\textrm{1Hot}}=\sqrt{3N}\,\delta_{i,c}.
\label{eq:1hot}
\end{equation}

To verify the correctness of the 1-hot stochastic estimator, we must check Eq. (\ref{eq:unitvar}). Since $v_{i}^{\textrm{1Hot}}$ has a single nonzero element, it is guaranteed

\begin{equation}
\left\langle v_{i}^{\textrm{\textrm{1Hot}}}v_{j}^{\textrm{\textrm{1Hot}}}\right\rangle =0\quad(i\neq j).
\end{equation}
Furthermore, the prefactor $\sqrt{3N}$ is precisely what is needed to satisfy
\begin{align*}
\left\langle v_{i}^{\textrm{\textrm{1Hot}}}v_{i}^{\textrm{\textrm{1Hot}}}\right\rangle  & =\frac{1}{3N}\sum_{c=1}^{3N}\left[v_{i}^{\textrm{\textrm{1Hot}}}\right]^{2}\\
 & =\frac{1}{3N}\sum_{c=1}^{3N}\left[\sqrt{3N}\,\delta_{i,c}\right]^{2}\\
 & =\sum_{c=1}^{3N}\delta_{i,c}\\
 & =1.
\end{align*}
These two equations demonstrate consistency with Eq. (\ref{eq:unitvar}). Therefore, both the Hutchinson random vectors $v^{\textrm{Hutch}}$ and the one-hot random vectors $v^{\textrm{\textrm{1Hot}}}$ yield valid and unbiased stochastic estimators. A detailed analysis of the stochastic error of the estimators used in this work is provided in the Supplementary Information and suggests that, under the physically motivated locality assumption that Hessian errors decay with interatomic distance, Hutchinson probing yields a mean-squared error that scales as $O(N)$ (implying an RMSE $\sim \sqrt{N}$ and a relative error that decreases with system size), whereas one-hot (column) probing concentrates error on diagonal terms and exhibits less favorable scaling, with the gap between the two estimators expected to widen as the number of atoms $N$ increases. The stochastic estimator based on $v^{\textrm{Hutch}}$ will be defined as the PHL method from now on.

\section{Methodology}

We investigate how second-derivative information can be efficiently incorporated into MLIP training. Starting from a chemically diverse quantum-chemical dataset, we train neural-network potentials under multiple supervision strategies that progressively include higher-order derivatives of the potential energy surface. Specifically, we compare standard energy–force (E-F) training with approaches that incorporate curvature information either through full Hessians or through HVPs. By keeping the model architecture and the underlying data fixed while varying only the form of derivative supervision, we isolate the impact of curvature information on training dynamics, predictive accuracy, and computational cost.

\subsection{Dataset Preparation}

To evaluate the role of Hessian information in training MLIPs, we employed the same dataset introduced in our previous work,\cite{rodriguez2025hessian,rodriguez2025data} which spans both equilibrium and non-equilibrium regions of a potential energy surface. All reference calculations were performed with Gaussian16 at the $\omega$B97XD/6-31G(d) level of theory.\cite{Gaussian16} Analytical Hessians were obtained through frequency analyzes. The choice of $\omega$B97XD provides a reliable treatment of non-covalent interactions and barrier heights, while the 6-31G(d) basis set balances accuracy and computational efficiency and ensures compatibility with the ANI-1x models.\cite{Smith2017-1,Smith2017-2,Smith2018,Smith2020,Gao2020,Devereux2020}

The dataset is divided into three complementary components:

\begin{enumerate}
    \item \textbf{Benchmark Test Set:} A collection of 35,087 equilibrium geometries from 11,961 reactions, including optimized reactants, products, and transition states. These structures were excluded from training and serve as a reference for interpolation accuracy.  
    \item \textbf{Intrinsic Reaction Coordinate (IRC) Dataset:} A set of 34,248 geometries sampled from 600 IRC trajectories, providing continuous reaction pathways that probe the curvature of the potential energy surface near transition states.  
    \item \textbf{Normal Mode Sampling (NMS) Dataset:} A collection of 62,527 non-equilibrium geometries generated by perturbing intermediate IRC structures along vibrational normal modes for 574 reactions. This dataset is designed to rigorously test the robustness of the extrapolation under large structural distortions.  
\end{enumerate}

The molecular systems in these datasets are \emph{small-molecule} geometries, with a typical atom count $N$ in the \emph{tens} (e.g., median $N\approx 14$ across the datasets) rather than the hundreds or thousands characteristic of condensed-phase simulations. This point is important because our asymptotic stochastic-error analysis is derived in the large-$N$ limit: as $N$ increases and Hessian errors remain localized, Hutchinson probing is predicted to become increasingly favorable relative to one-hot (one-column) probing. Thus, while the present benchmarks already show clear advantages for Hutchinson probing in data-limited settings, the theoretical scaling suggests that these advantages should grow for larger systems such as extended materials, large clusters, or supercells.

For our purposes, these datasets provide diverse coverage of the potential energy surface, with the benchmark test set and IRC dataset assessing interpolation performance in chemically relevant regions, and the NMS dataset challenging the models under far-from-equilibrium conditions. These datasets form the common foundation for all the training strategies described below, enabling direct comparison of how different forms of derivative supervision influence learning behavior and generalization.

\subsection{Training Objective and Strategies}

Using these datasets, our objective was to quantify the trade-off between predictive accuracy and computational efficiency as progressively higher-order derivative information is introduced into MLIP training. To this end, we evaluated four training schemes that progressively incorporate higher-order derivatives of the potential energy surface into the objective function. The schemes are designed to explore the tradeoffs between accuracy and computational cost; a detailed description of the optimization setup (dataset split, hyperparameters, batching, probe sampling, and other information) is provided in the Supplementary Information under \emph{Training Procedure}. The training schemes are the following:

\begin{enumerate}
    \item \textbf{E--F:} Training on total energies $E$ and atomic forces $F$. This baseline reflects standard practice in the development of MLIPs.
    \item \textbf{E--F--HVP (one-column):} Training on energies, forces, and Hessian--vector products (HVP), where $Hv = \nabla^2 E(\mathbf{R})\,v$ is evaluated with a single one-hot probe vector $v^{\mathrm{1Hot}}$ aligned with a coordinate axis. This approach introduces curvature information at minimal additional cost and has been recently used for training MLIPs to Hessian information.\cite{cui2025large,koker2026pft}
    \item \textbf{E--F--HVP (PHL):} Training on energies, forces, and HVPs using \emph{Projected Hessian Learning}, where $Hv = \nabla^2 E(\mathbf{R})\,v$ is evaluated using Hutchinson-type Gaussian probing vectors $v^{\mathrm{Hutch}}$ with mean zero and unit variance. This stochastic estimator provides an unbiased approximation to full Hessians.
    \item \textbf{E--F--H:} Training on energies, forces, and full Hessians. This provides the most complete curvature information but is associated with a prohibitive increase in computational cost.
\end{enumerate}

In summary, E-F-H represents the upper bound in accuracy but is computationally expensive, while E-F serves as the lowest-cost baseline. The two HVP-based methods offer intermediate strategies that capture much of the benefit of Hessian training while remaining computationally efficient.

In both HVP-based schemes, we considered two vector protocols: (i) \emph{randomized-vector} training, in which new probe vectors are resampled at every minibatch, and (ii) \emph{fixed-vector} training, in which a single probe vector per configuration is retained throughout training. The fixed-vector setting serves as a controlled surrogate for scenarios in which the ground-truth Hessian is not known in full, but itself sampled.

Finally, we examine both the running time of the machine-learning model training and the cost of generating the underlying quantum mechanical reference data to assess the efficiency and benefits of the different training schemes. We compared the cost of computing energies, forces, full Hessians and HVPs using DFT. To allow fair comparison between training schemes, all models were optimized using a unified loss framework that differs only in how curvature information is incorporated.

\subsection{Loss Function Formulation}
The total training loss was defined as a weighted sum of errors in predicted energies, forces, and, when included, Hessian information. Two formulations were used depending on whether the model was trained with full Hessians or with Hessian–vector products:  
\begin{align}
\mathcal{L}_{\text{full}} & =
\mathcal{L}_{E}(E^{\text{pred}}, E^{\text{ref}})
+ \lambda_{F}\,\mathcal{L}_{F}(F^{\text{pred}}, F^{\text{ref}})
+ \lambda_{H}\,\mathcal{L}_{H}(H^{\text{pred}}, H^{\text{ref}}), \\
\mathcal{L}_{\text{HVP}} & =
\mathcal{L}_{E}(E^{\text{pred}}, E^{\text{ref}})
+ \lambda_{F}\,\mathcal{L}_{F}(F^{\text{pred}}, F^{\text{ref}})
+ \lambda_{H}\,\hat{\mathcal{L}}_{H}(H^{\text{pred}} v, H^{\text{ref}} v).
\end{align}

Here, \(E\) represents the total molecular energies, \(F\) represents atomic forces, \(H\) represents the Hessian matrices, and \(v\) represents probe vectors. In the full-Hessian formulation, the model is trained against all second derivatives, whereas in the HVP formulation, only products of the Hessian with probe vectors are included. The weights ($\lambda_{F}=0.30$) and ($\lambda_{H}=0.09$) were tuned to balance the relative contributions of forces and Hessian information against energies. Typically, energies set the absolute scale, while forces and Hessian terms are given smaller weights to ensure they contribute comparably during optimization. All models were trained under identical optimization settings, with differences arising solely from the form of derivative supervision.

\section{Results and Discussion}

\subsection{Validation Loss Convergence}

\begin{figure}[ht]
    \centering
        \includegraphics[width=\textwidth]{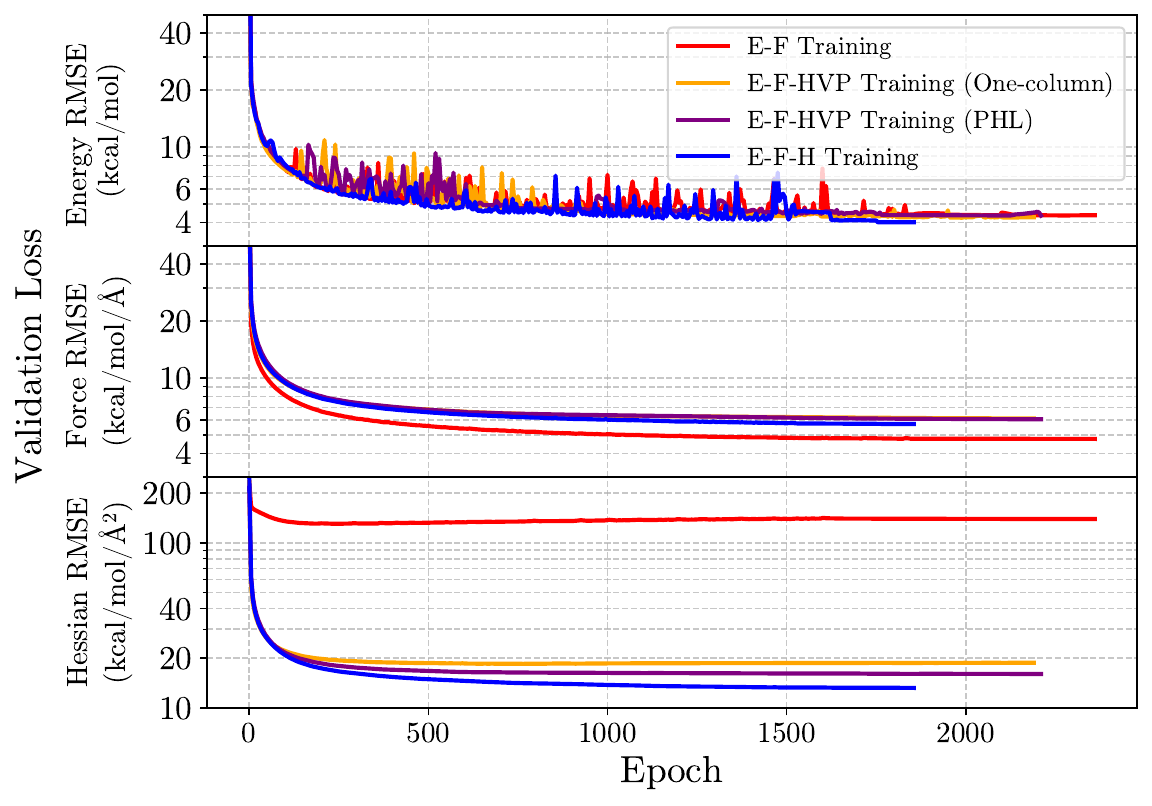}
    \caption{Validation loss curves for energy, force, and Hessian predictions under the fixed-vector approach. Each panel shows the RMSE of the validation set as a function of training epoch for energy and force training (E–F, red), energy, force, and Hessian–vector product training using the one-column method (E–F–HVP One-Column, orange), energy, force, and HVP training using the Hutchinson estimator (E–F–HVP PHL, purple), and full energy–force–Hessian training (E–F–H, blue).
}
    \label{fig:training_RMSE_fixed}
\end{figure}

The convergence behavior of the different training schemes was examined using validation RMSE curves for energies, forces, and Hessians under the fixed-vector approach (Figure~\ref{fig:training_RMSE_fixed}). These curves provide insight into how curvature information influences optimization dynamics, training stability, and the relative behavior of one-column and Hutchinson-based Hessian-vector product estimators (PHL). The validation results for the randomized-vector experiments reached comparable final RMSE values for the characteristic system size of our training dataset (a median of 14 atoms per system) and are therefore omitted from this section for brevity; their performance in generalization is analyzed separately below.

For energy errors, all training schemes eventually converge to similar RMSE validation levels. However, clear differences appear during the early and intermediate stages of training. Methods incorporating curvature information exhibit reduced fluctuations in validation loss after approximately 700 epochs, resulting in smoother and more stable convergence trajectories. Full Hessian training (E–F–H) shows the most stable behavior early in training and converges more rapidly, while both HVP-based approaches achieve greater stability than the E–F baseline during the intermediate and late stages of training.

This smoother convergence suggests that the incorporation of second-derivative information regularizes the optimization process, reducing the sensitivity to stochastic noise and sharp variations in the loss landscape. Such stabilization effects have been widely observed in both machine learning and atomistic modeling,\cite{hardt2016train,hochreiter1997flat} where smoother optimization trajectories are often associated with improved generalization and robustness, even when the final training errors are similar.

For force errors, E–F training achieves the lowest absolute validation RMSE, consistent with its objective function placing the greatest emphasis on force accuracy. Importantly, introducing second-derivative information, either via full Hessians or HVP estimators, does not degrade force prediction accuracy. Both E-F-HVP models closely track the convergence behavior of E–F–H, indicating that curvature information can be incorporated without compromising force fidelity. In particular, although E–F achieves slightly lower force RMSE in the validation set, this does not necessarily translate to improved generalization, particularly for configurations outside of the training distribution, as discussed in the following section.

For Hessian errors, the benefit of incorporating second-derivative information is most pronounced under the data regimes considered here (i.e., equilibrium and saddle-point structures). Models trained only on energies and forces exhibit large and poorly converged Hessian validation errors, indicating that curvature information is not reliably recovered from first-order data alone. In contrast, compared to the E-F training, incorporating curvature supervision through HVPs reduced Hessian RMSE by 71-86\% for the one-column estimator and 74-88\% for the Hutchinson estimator across the Test, IRC, and NMS datasets. The full Hessian training unsurprisingly achieves the lowest overall error with 77-90\% RMSE reduction compared to the E-F training. It is important to note that the Hutchinson estimator achieved 11–15\% lower Hessian RMSE than the one-column estimator in the fixed-vector training scheme, even for our "small" system sizes.

Crucially, HVP-based methods recover much of the curvature accuracy and training stability of full Hessian supervision at a dramatically reduced computational cost, making them effective in regimes where explicit Hessian evaluation is cost prohibitive. Overall, HVP-based training captures most of the optimization and generalization benefits of full Hessian inclusion, with the Hutchinson-based PHL estimator offering improved robustness and accuracy when only a single fixed HVP per molecular system is available.

\subsection{Predictive Accuracy Across Datasets}

To evaluate generalization performance, we compared the final RMSE values of all models on the Test Set, the IRC dataset, and the NMS dataset. Figure \ref{fig:rmse_bars} summarizes the results under the randomized-vector and fixed-vector approaches with separate panels for the energy, force, and Hessian predictions.

\subsubsection{Randomized HVP Vectors per Minibatch}

\begin{figure}[htbp!]
    \centering
    \begin{subfigure}[b]{0.44\textwidth}
        \centering
        \includegraphics[width=\textwidth]{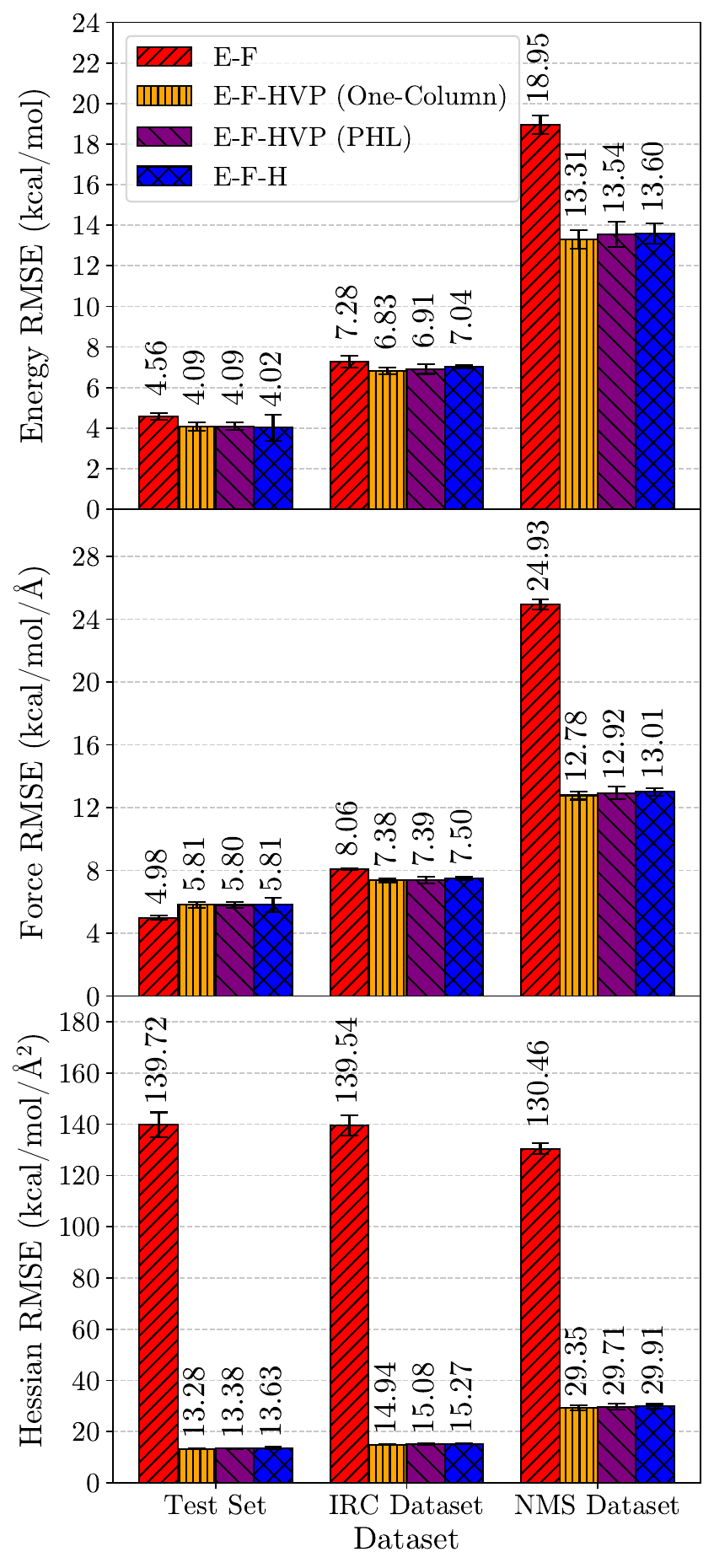}
        \caption{Randomized vectors}
        \label{subfig:rmse_randomized}
    \end{subfigure}
    \hfill
    ~~
    \begin{subfigure}[b]{0.44\textwidth}
        \centering
        \includegraphics[width=\textwidth]{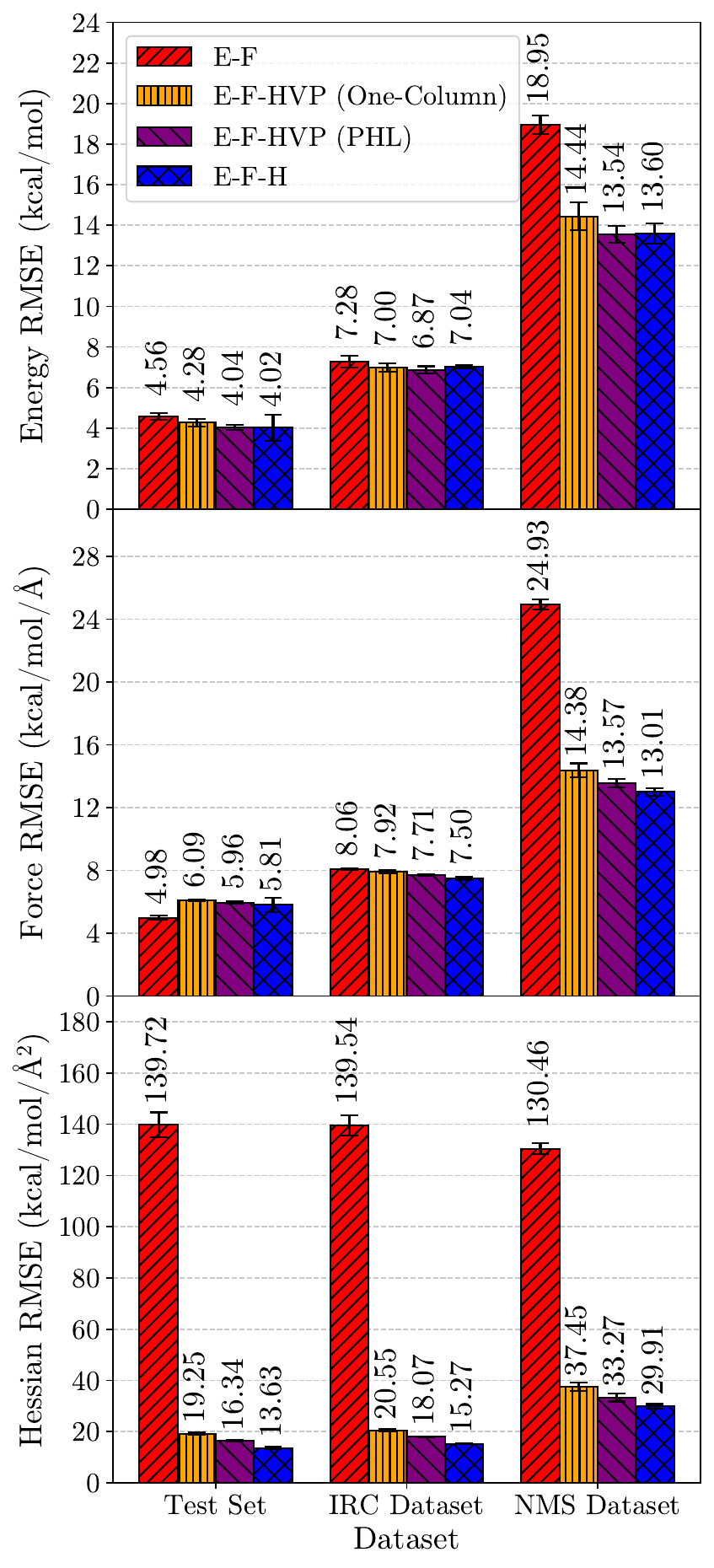}
        \caption{Fixed vectors}
        \label{subfig:rmse_fixed}
    \end{subfigure}
    \caption{Validation RMSE comparison across the Test Set, IRC dataset, and NMS dataset for models trained with different levels of information using (a) \emph{randomized} probe vectors (resampled each minibatch) and (b) \emph{fixed} probe vectors (one probe per system). Results are shown for energy and force training (E–F); energy, force, and Hessian–vector products (HVP) using the one-column method (E–F–HVP One-Column); energy, force, and HVP using the Hutchinson estimator (E–F–HVP PHL); and energy, force, and full Hessian training (E–F–H). From top to bottom, panels report RMSE for energies, forces, and Hessians. Bars represent mean RMSE values, with error bars indicating variability across ensembles of five independently trained models. Incorporating Hessian information improves accuracy relative to E-F across all datasets; under randomized probing, one-hot and PHL achieve statistically indistinguishable performance for the small molecular systems studied here, whereas under fixed probes PHL provides systematically lower errors, most notably for extrapolative NMS geometries.}

    \label{fig:rmse_bars}
\end{figure}

When the HVP probe vector is randomized in each minibatch step (Figure \ref{subfig:rmse_randomized}), the performance of the one-column and PHL estimators is nearly indistinguishable across all datasets and all training targets for the system sizes in the training data.

\begin{itemize}
    \item \textbf{Energy RMSE:} Under randomized-vector training, both HVP estimators substantially improve upon the E–F baseline and approach full Hessian accuracy. On the NMS dataset, the energy RMSE is reduced by approximately 29\% compared to E–F for both one-column and PHL estimators, with only a difference of 1 to 2\% between them. Similar trends are observed on the Test and IRC datasets, indicating that stochastic curvature sampling provides sufficient coverage for accurate energy prediction.
    \item \textbf{Force RMSE:} Although E–F remains marginally optimal for in-domain force prediction, HVP-based methods closely match full Hessian performance and substantially outperform E–F on extrapolative NMS geometries. Specifically, the force RMSE on NMS is reduced by approximately 48 to 49\% relative to E–F for both HVP estimators, with a separation of less than 2\% between the one-column and PHL approaches.
    \item \textbf{Hessian RMSE:} HVP training dramatically improves Hessian prediction accuracy compared to E–F, reducing the NMS Hessian RMSE by approximately 77\% for both estimators. The one-column and PHL methods exhibit nearly identical performance in this randomized setting, differing only by $\sim$1\%, confirming that minibatch-level randomization provides sufficiently rich curvature sampling to approximate full Hessian supervision.
\end{itemize}
Randomized HVP vectors therefore enable broad curvature coverage during training, yielding performance nearly indistinguishable from full Hessian training for energies, forces, and Hessians, while avoiding the substantial computational cost of explicit Hessian supervision. Note also that our theoretical analysis suggests that the Hutchinson-based PHL method would become preferred when training on larger system sizes $N$, for which the $N \times N$ Hessian matrix carries much more information.

\subsubsection{Fixed HVP Vectors per System}

The fixed-vector setting (Figure~\ref{subfig:rmse_fixed}) provides a more realistic scenario in which only one HVP per system is available from quantum chemistry. Under these conditions, the differences between the one-column and PHL estimators become more pronounced.

\begin{itemize}
    \item \textbf{Energy RMSE:} All Hessian-informed methods improve energy RMSE relative to E-F training across all datasets, with gains that are largest on extrapolative geometries. On the Test set, energy RMSE decreases by 6.1\% with the one-column estimator and 10.3\% with PHL, compared to E-F (11.8\% for full Hessian training). On the IRC dataset, the reductions are smaller but consistent: 3.8\% for one-column and 5.6\% for PHL. On the NMS dataset, the improvements are substantial: one-column reduces the energy RMSE by 23.8\% and PHL by 28.5\% relative to E-F, with PHL achieving a 6.2\% reduction in RMSE compared to one-column (and essentially matching complete Hessian training, 28.2\%).
    \item \textbf{Force RMSE:} The benefit of curvature supervision is most apparent for intermediate or NMS configurations. Although E-F training attains the lowest RMSE on the Test set, both HVP methods match force accuracy on the IRC geometries and substantially improve force accuracy on NMS geometries. Compared to E-F, the one-column and PHL estimators reduce the force RMSE of the NMS dataset by 42.3\% and 45.6\%, respectively (47.8\% for full Hessian training). The PHL estimator consistently achieves a lower RMSE than the one-column estimator across datasets, with a 2.1\% reduction in RMSE on the Test set, 2.7\% on the IRC set, and 5.6\% on the NMS set.
    \item \textbf{Hessian RMSE:} The strongest separation between estimators appears for Hessian prediction. Relative to E-F training, Hessian RMSE is reduced by 86.2\% with one-column estimators and 88.3\% with PHL estimators on the Test set; 85.3\% and 87.0\% on IRC; and 71.3\% and 74.5\% on NMS. The PHL estimator consistently outperforms the one-column estimator across all datasets, with Hessian RMSE lower by 15.1\% on the Test set, 12.1\% on IRC, and 11.2\% on NMS.
\end{itemize}

When only a single HVP per molecular configuration is available, the PHL estimator yields errors that are systematically lower than those of the one-column approach across all predicted properties. On the extrapolative NMS dataset, PHL reduces the energy RMSE by 6.2\%, the force RMSE by 5.6\%, and the Hessian RMSE by 11.2\% relative to the one-column estimator. These improvements indicate that random Hutchinson vectors provide a more uniform sampling of curvature directions under fixed-vector constraints. While the one-column estimator remains a computationally inexpensive alternative, its reliance on a single coordinate direction limits performance in data-sparse regimes and large system sizes.

\subsection{Statistical Significance of RMSE Differences}

To quantify whether the one-column and PHL estimators produce statistically distinguishable predictive accuracy, we conducted paired two-sample $t$-tests using the RMSE values from the five independently trained models for each method. The tests were performed separately for the randomized-vector and fixed-vector regimes, allowing us to isolate the effect of the vector-sampling strategy on estimator behavior. The results are summarized in Tables \ref{tab:t_test_results_randomized} and \ref{tab:t_test_results_fixed}, where the mean RMSE difference is defined as (one-column RMSE – PHL RMSE). Thus, negative values indicate lower RMSE for the one-column estimator, while positive values indicate lower RMSE for the PHL estimator.

\subsubsection{Randomized HVP Vectors per Minibatch}

Table \ref{tab:t_test_results_randomized} reports the results of the paired $t$-test for experiments in which the HVP vector was randomly resampled in each minibatch. Under this configuration, no statistically significant differences were observed between the two estimators across any of the datasets or properties used (all $p > 0.05$).

\begin{table}[ht!]
\centering
\caption{Two-sample $t$-test results comparing RMSE values from the one-column and PHL estimators under the randomized-vector training approach. 
The mean difference is defined as (one-column – PHL). 
Negative values indicate lower RMSE for the one-column method. 
Asterisks denote statistically significant differences ($p<0.05$).
}
\label{tab:t_test_results_randomized}
\begin{tabular}{lcccc}
\hline
\textbf{Property} & \textbf{Dataset} & \textbf{Mean Difference (kcal/mol)} & \textbf{$p$-value} & \textbf{Significance} \\
\hline
Energy  & Test & -0.0039 & 0.950 & – \\
Energy  & IRC  & -0.0775 & 0.348 & – \\
Energy  & NMS  & -0.2320 & 0.342 & – \\
\hline
Force   & Test & 0.0134 & 0.832 & – \\
Force   & IRC  & -0.0198 & 0.732 & – \\
Force   & NMS  & -0.1444 & 0.195 & – \\
\hline
Hessian & Test & -0.1051 & 0.405 & – \\
Hessian & IRC  & -0.1468 & 0.166 & – \\
Hessian & NMS  & -0.3608 & 0.189 & – \\
\hline
\end{tabular}
\end{table}

Across energies, forces, and Hessians, the mean RMSE differences are small and fluctuate around zero, consistent with the interpretation that randomized stochastic sampling effectively averages out directional bias for our characteristic system sizes. Randomized vectors ensure broad coverage of curvature directions throughout training, allowing both estimators to recover similarly accurate approximations to the Hessian.

These results indicate that when HVP vectors vary throughout training, the one-column and PHL estimators perform equivalently, and neither method shows a statistically significant advantage for these system sizes.

\subsubsection{Fixed HVP Vectors per System}

The corresponding results of the paired $t$-test for fixed-vector experiments are shown in Table \ref{tab:t_test_results_fixed}. In the fixed-vector regime, a single HVP vector is assigned to each molecule and reused across all epochs. This setting simulates scenarios in which only one HVP is available per system from quantum chemical calculations. Under this more restrictive condition, statistically significant differences emerge between the two estimators for several datasets and training targets.

\begin{itemize}
    \item \textbf{Energy RMSE:} For the Test and IRC datasets, the mean differences are not statistically significant ($p > 0.05$), indicating comparable energy predictions. However, for the NMS dataset, the PHL estimator yields significantly lower energy RMSE ($p = 0.006$), demonstrating better extrapolative capability.
    \item \textbf{Force RMSE:} The PHL estimator performs significantly better on the IRC dataset ($p = 0.013$) and the NMS dataset ($p = 0.006$). This suggests that its broader curvature sampling improves first-derivative accuracy in intermediate and far-from-equilibrium geometries, where directional bias from a single one-column vector becomes limiting.
    \item \textbf{Hessian RMSE:} Across all datasets, the PHL estimator yields a significantly lower Hessian RMSE ($p < 0.01$ for all comparisons). These results confirm that the Hutchinson vectors provide substantially better curvature information when only one vector per molecule is available.
\end{itemize}

\begin{table}[ht]
\centering
\caption{Two-sample $t$-test results comparing RMSE values from the one-column and PHL estimators under the fixed-vector training approach. 
The mean difference is defined as (one-column – PHL). 
Negative values indicate lower RMSE for the one-column method. 
Asterisks denote statistically significant differences ($p<0.05$).
}
\label{tab:t_test_results_fixed}
\begin{tabular}{lcccc}
\hline
\textbf{Property} & \textbf{Dataset} & \textbf{Mean Difference (kcal/mol)} & \textbf{$p$-value} & \textbf{Significance} \\
\hline
Energy  & Test & 0.2370 & 0.063 & – \\
Energy  & IRC  & 0.1303 & 0.399 & – \\
Energy  & NMS  & 0.8989 & 0.006 & * \\
\hline
Force   & Test & 0.1268 & 0.097 & – \\
Force   & IRC  & 0.2076 & 0.013 & * \\
Force   & NMS  & 0.8086 & 0.006 & * \\
\hline
Hessian & Test & 2.9105 & 0.0000029 & * \\
Hessian & IRC  & 2.4722 & 0.00022 & * \\
Hessian & NMS  & 4.1780 & 0.005 & * \\
\hline
\end{tabular}
\end{table}

This analysis indicates that when curvature information is limited to a single HVP per system, the PHL estimator consistently outperforms the one-column method, especially for the NMS dataset and for Hessian predictions, due to its isotropic distribution and reduced directional bias.

\subsection{Computational Efficiency}

A key motivation for incorporating Hessian–vector products into MLIP training is the ability to reap the benefits of second-derivative information without incurring the cost of full Hessian evaluations. To quantify this advantage, we assessed computational efficiency at two levels: (1) MLIP training runtime per epoch in Figure~\ref{fig:training_speedup} and (2) quantum-chemistry cost scaling for computing energies, forces, Hessians, and HVPs in Figure~\ref{fig:cost_scaling}.

\subsubsection{MLIP Training Runtime}

\begin{figure}[htbp]
    \centering
    \begin{subfigure}[b]{0.48\textwidth}
        \centering
        \includegraphics[width=\textwidth]{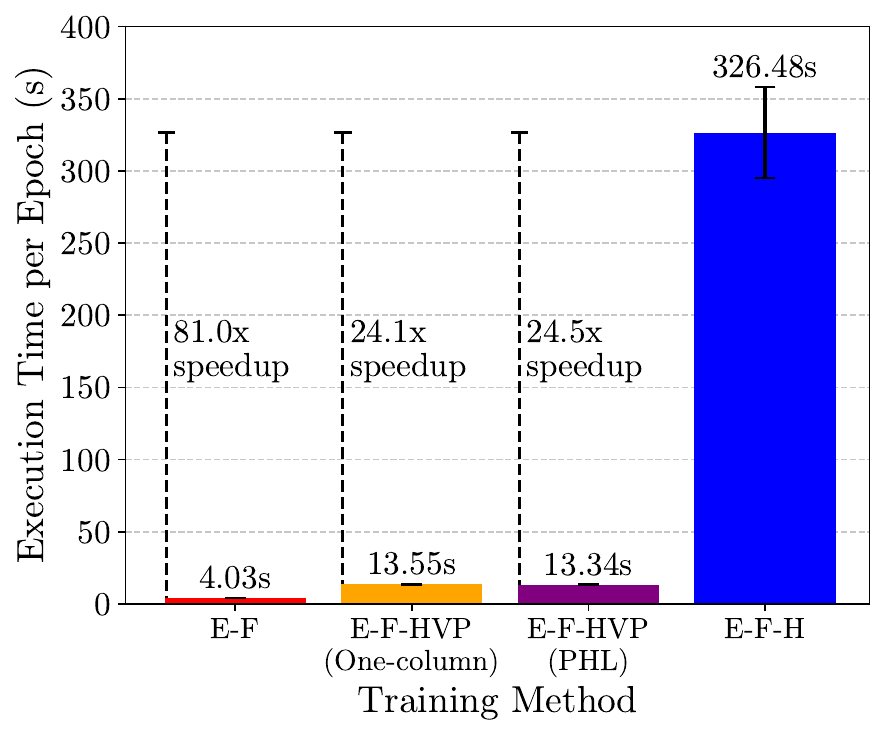}
        \label{fig:training_cost}
    \end{subfigure}
    \hfill
    \begin{subfigure}[b]{0.48\textwidth}
        \centering
        \includegraphics[width=\textwidth]{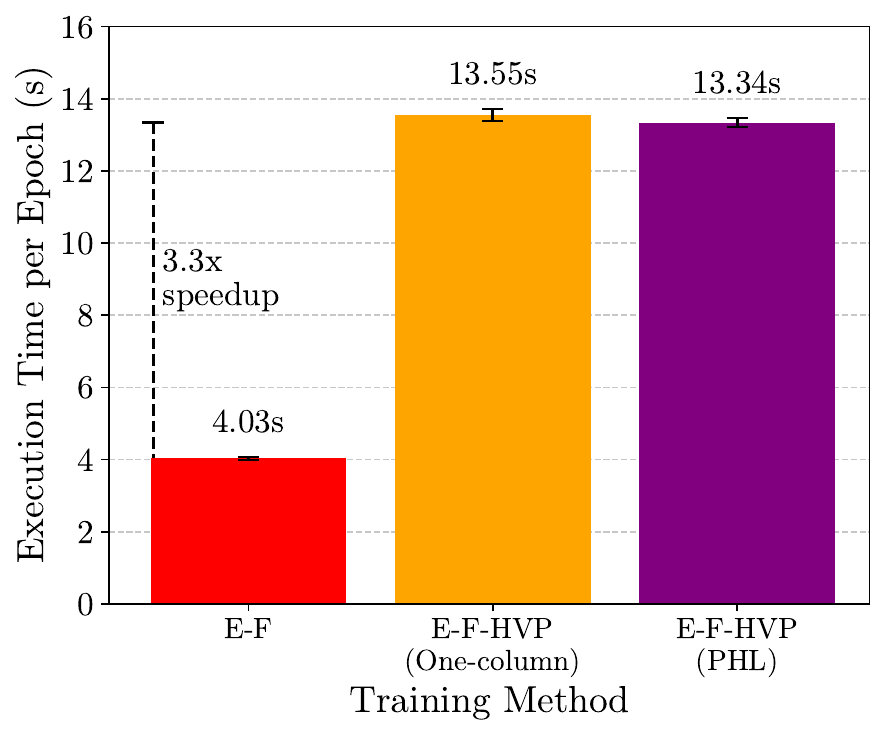}
        \label{fig:training_cost_zoom}
    \end{subfigure}
    \caption{Execution time per training epoch for different methods: energy and force training (E–F); energy, force, and Hessian–vector products (HVP) using the one-column method (E–F–HVP One-Column); energy, force, and HVP using the PHL method (E–F–HVP PHL); and energy, force, and full Hessian training (E–F–H). Bars show mean times with error bars indicating variability across epochs. Both one-column and PHL estimators achieve more than a 24-fold speedup compared to full Hessian training, while E–F training is the fastest but lacks Hessian information.
}
    \label{fig:training_speedup}
\end{figure}

Figure \ref{fig:training_speedup} reports the execution time per epoch for all training schemes. Full Hessian training (E–F–H) is by far the most computationally demanding, requiring over an order of magnitude more time per epoch than the other methods. This cost arises from the need to backpropagate through the complete Hessian matrix, which involves repeated second-derivative evaluations.

In contrast, both HVP-based approaches (E–F–HVP One-Column and E–F–HVP PHL) achieve epoch times of approximately 13 s, representing a 24-fold speedup relative to full Hessian training. Importantly, both estimators preserve most of the predictive gains afforded by full Hessians, providing a near-optimal tradeoff between speed and accuracy.

Energy–force training (E–F) remains the fastest configuration at roughly 4 s per epoch, but lacks any curvature information. The HVP methods therefore occupy a favorable middle ground: only three times slower than E–F training but dramatically more informative. This comparison confirms that incorporating curvature information via HVPs is highly practical in large-scale MLIP training pipelines, particularly when full Hessians would be computationally prohibitive.

\subsubsection{Quantum-Chemistry Cost Scaling}

\begin{figure}[ht]
    \centering
    \begin{subfigure}[b]{0.48\textwidth}
        \centering
        \includegraphics[width=\textwidth]{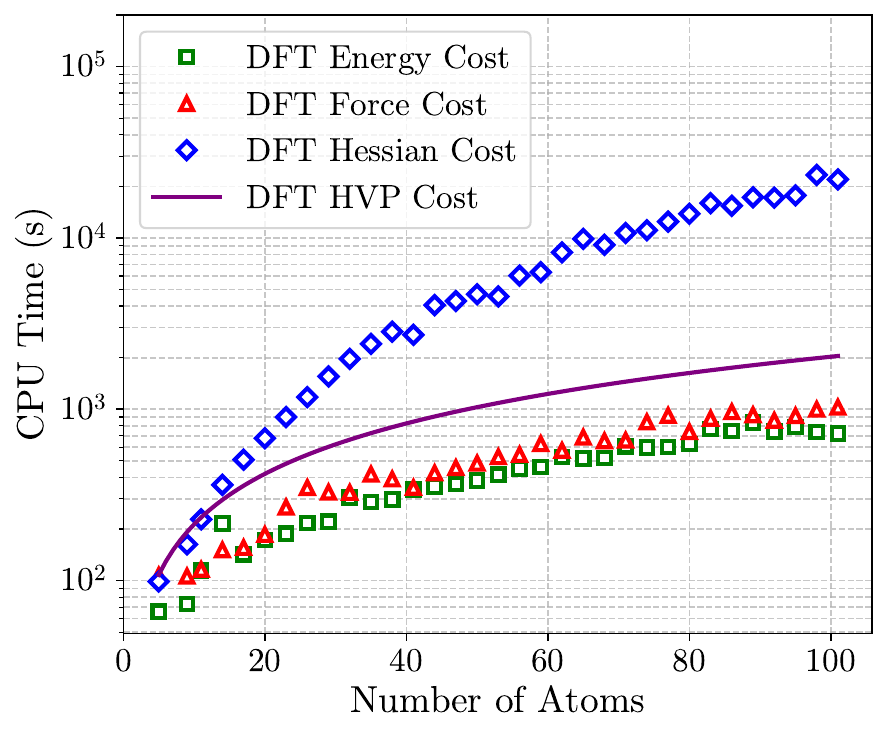}
        \caption{Semi-log plot}
        \label{fig:cost_scaling_log}
    \end{subfigure}
    \hfill
    \begin{subfigure}[b]{0.48\textwidth}
        \centering
        \includegraphics[width=\textwidth]{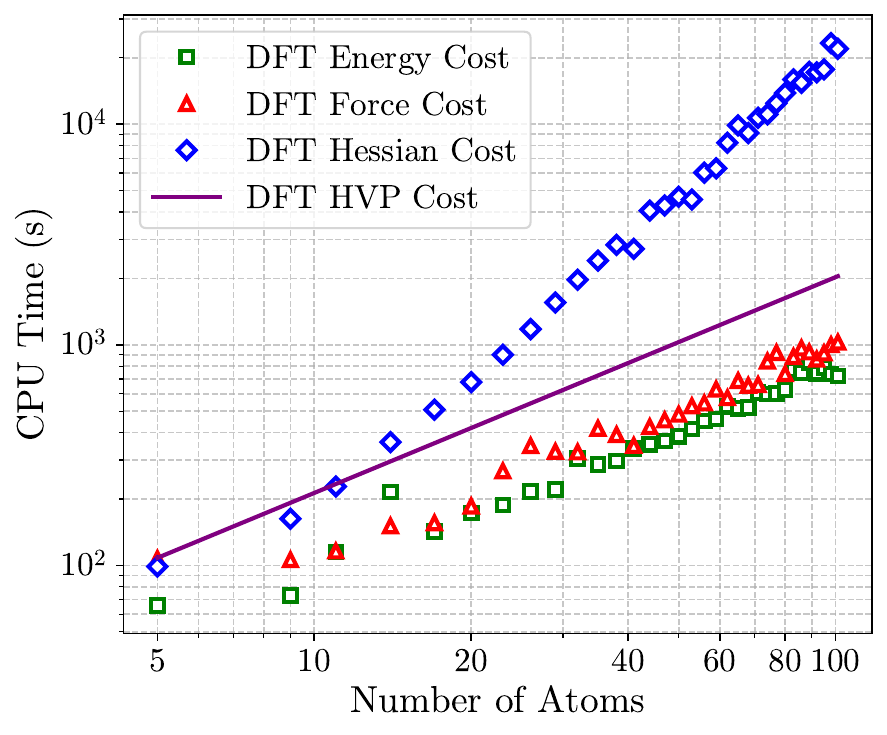}
        \caption{Log-log plot}
        \label{fig:cost_scaling_log-log}
    \end{subfigure}
    \caption{Scaling of CPU time with system size for density functional theory (DFT) calculations using Gaussian16 on a semi-log plot (a) and a log-log plot (b). Shown are the computational costs for evaluating total energies (green squares), forces (red triangles), and full Hessians (blue diamonds), and the estimated cost for evaluating Hessian–vector products (HVP, purple line). While force and energy costs scale comparably, the cost of full Hessian evaluations grows much more steeply with the number of atoms. In contrast, HVP evaluations provide a significantly cheaper alternative, with scaling closer to that of two force calculations, highlighting their efficiency for incorporating second-derivative information.
}
    \label{fig:cost_scaling}
\end{figure}

To contextualize the MLIP results with first-principles cost considerations, Figure \ref{fig:cost_scaling} presents the CPU scaling for evaluating energies, forces, full Hessians, and approximate HVP costs using Gaussian16. Energy and force calculations show similar scaling behavior as system size increases, reflecting their reliance on first derivatives of the electronic energy. However, full Hessian evaluations grow much more steeply. This superlinear scaling rapidly dominates computational budgets for systems beyond a few dozen atoms, limiting the practical generation of high-quality second-derivative datasets. 

By contrast, HVP evaluations scale much closer to forces. A single Hessian-vector product can be computed using two force-like operations via forward-over-backward or backward-over-backward automatic differentiation. Alternatively, when the probing direction $v$ is fixed, the same quantity can be obtained from finite differences of forces by evaluating $\mathbf{F}(\mathbf{R}\pm\epsilon v)$ and forming
$
Hv \approx -\big[\mathbf{F}(\mathbf{R}+\epsilon v)-\mathbf{F}(\mathbf{R}-\epsilon v)\big]/(2\epsilon),
$
which again requires only two force evaluations. As a result, a single HVP is expected to cost on the order of two force evaluations, making it orders of magnitude cheaper than computing a full Hessian matrix. Because HVPs capture directional curvature information at such a low cost, they are an attractive surrogate for full Hessians both at the quantum-chemistry level (for data generation) and at the MLIP training level (for incorporating curvature into the loss function).

\section{Conclusions}

We introduced \emph{Projected Hessian Learning} (PHL), a scalable curvature-supervision framework for MLIPs that replaces explicit Hessian construction with fast Hessian-vector product (HVP) probes. Using a chemically diverse dataset that spans reactants, products, transition states, intrinsic reaction coordinates (IRC), and normal-mode-sampled (NMS) geometries, we compared four training strategies of increasing physical fidelity: energy–force (E-F) training, HVP-based (E-F-HVP) training using one-column or PHL estimators, and full Hessian (E–F–H) training.

Incorporating curvature information stabilizes training dynamics and improves predictive accuracy. Across Test, IRC, and NMS datasets, the HVP-trained models substantially outperform the E-F baselines, approach full-Hessian training accuracy for the energies and forces, and reduce Hessian errors by approximately 77\% on extrapolative NMS geometries. Under randomized-vector training, both HVP-based estimators achieve nearly identical performance for our system sizes, reducing the NMS energy RMSE by $\sim$29\%, the force RMSE by $\sim$48\% and the Hessian RMSE by $\sim$77\% relative to E-F, indicating that stochastic curvature sampling effectively approximates full Hessian supervision. In contrast, under fixed-vector conditions representative of data-limited regimes, Hutchinson-based PHL consistently outperforms one-hot estimators, yielding additional reductions of 6.2\% in energy RMSE, 5.6\% in force RMSE, and 11.2\% in Hessian RMSE on NMS geometries, as confirmed by paired $t$-tests.

Computationally, full Hessian training is dramatically more expensive, requiring on average 326.5 s per epoch, compared to 13.6 s and 13.3 s for the one-column probing and our Hutchinson-based PHL, respectively, corresponding to $\sim$24$\times$ speedups relative to E–F–H while incurring only a modest overhead compared to standard E–F training. This improvement arises because full Hessian training requires evaluating and backpropagating through all second-derivative components $(3N)^2$, while HVP-based methods access curvature only through Hessian-vector products, whose cost scales similarly to force evaluations. As a result, HVP-based approaches recover most of the benefits of Hessian supervision at less than 8\% of the computational cost of full Hessian training. Consistent with this observation, at the quantum-chemistry level, HVP evaluations scale comparably to forces and far more favorably than full Hessians, making them a practical surrogate for incorporating second-derivative information in MLIP training.

These results establish PHL as an efficient and scalable alternative to full Hessian supervision, delivering substantial accuracy gains over energy-force models at a greatly reduced cost. The PHL perspective also naturally connects to recent \emph{phonon fine-tuning} (PFT) work, which uses stochastic curvature supervision to improve force-constant (phonon) accuracy in periodic materials; PHL provides a general projected-curvature framework that can be applied to materials settings where curvature governs vibrational and elastic response, and where large supercells make explicit Hessians impractical.

Future work will extend PHL to larger and more complex systems, develop adaptive probing strategies, and integrate projected-curvature supervision with active learning and uncertainty quantification. Our asymptotic analysis indicates that Hutchinson probing should become increasingly advantageous over one-hot probing as system size $N$ grows, suggesting particularly strong benefits in condensed-phase and materials settings beyond the tens-of-atoms regime explored here (typical $N=14$). In addition, when the probing vector $v$ is fixed, reference HVPs, i.e., $Hv$, can be computed in specialized quantum chemistry codes via finite differences of gradients along $v$ (forces in geometries displaced by $\pm\epsilon v$), enabling curvature-informed training on much larger systems and facilitating access to bulk materials properties that require large supercells, such as surfaces and defects.

\section*{Data and code availability}
The datasets used in this study are publicly available through the OpenREACT Figshare repository (OpenREACT-CHON-EFH; energies, forces, and Hessians for the RTP/IRC/NMS datasets used in this work): \url{https://doi.org/10.6084/m9.figshare.29189858}.

The code used to train and evaluate the ANI models in this paper, including notebooks and instructions for accessing the datasets, is available at: \url{https://github.com/Austinrg14/PHL}.

In addition, we implemented Hessian and Hessian--vector product (HVP) training functionality in the HIPPYNN framework to enable curvature-supervised training of HIP-NN models (including full-Hessian and projected-curvature/PHL training workflows). 

HIPPYNN is available through the Los Alamos National Laboratory (LANL) GitHub repository: \url{https://github.com/lanl/hippynn}. An example script demonstrating our implementation is provided at \texttt{hippynn/examples/hessian\_training.py}.

\section*{Acknowledgements}
This work is supported by the U.S. Department of Energy, Office of Basic Energy Sciences (FWP: LANLE8AN and FWP: LANLE3F2) and by the U.S. Department of Energy through Los Alamos National Laboratory. Los Alamos National Laboratory is operated by Triad National Security, LLC, for the National Nuclear Security Administration of the U.S. Department of Energy Contract No. 892333218NCA000001. This research used resources provided by the LANL CAI-1 Darwin cluster at LANL.  This work was supported in part through computational resources and services provided by the Institute for Cyber-Enabled Research at Michigan State University (MSU). This work used the MSU Data Machine, which is supported through the NSF Campus CyberInfrastructure program through grant \#2200792.

\bibliographystyle{unsrtnat}
\bibliography{references}

@article{yuan2024analytical,
  author = {{Yuan, Eric C.-Y. and Kumar, Anup and Guan, Xingyi and Hermes, Eric D. and Rosen, Andrew S and Z{\'a}dor, Judit and Head-Gordon, Teresa and Blau, Samuel M.}},
  title = {Analytical ab initio Hessian from a deep learning potential for transition state optimization},
  journal = {Nature Communications},
  year = {2024},
  volume = {15},
  number = {1},
  pages = {8865},
  publisher = {Nature Publishing Group UK London},
  doi = {10.1038/s41467-024-52481-5}
}

@article{rodriguez2025hessian,
  author = {Rodriguez, Austin and Smith, Justin S and Mendoza-Cortes, Jose L},
  title = {Does Hessian Data Improve the Performance of Machine Learning Potentials?},
  journal = {Journal of Chemical Theory and Computation},
  year = {2025},
  volume = {21},
  number = {14},
  pages = {6698--6710},
  publisher = {ACS Publications},
  doi = {10.1021/acs.jctc.5c00402}
}

@misc{rodriguez2025data,
  author = {Rodriguez, Austin and Smith, Justin S. and Mendoza-Cortes, Jose L.},
  title = {{OpenREACT-CHON-EFH — Open REaction Dataset of Atomic ConfiguraTions comprising C, H, O, N with Energies, Forces, and Hessians}},
  year = {2025},
  doi = {10.6084/m9.figshare.29189858},
  month = {5},
  note = {Figshare dataset, doi:10.6084/m9.figshare.29189858}
}

@article{cui2025large,
  author = {Cui, Taoyong and Han, Yonghong and Jia, Haojun and Duan, Chenru and Zhao, Qiyuan},
  title = {A Large Scale Molecular Hessian Database for Optimizing Reactive Machine Learning Interatomic Potentials},
  journal = {Scientific Data},
  year = {2026},
  volume = {13},
  number = {1},
  pages = {37},
  publisher = {Nature Publishing Group UK London},
  doi = {10.1038/s41597-025-06350-5}
}

@article{williams2025hessian,
  author = {Williams, Nicholas J and Kabalan, Lara and Stojanovic, Ljiljana and Z{\'o}lyomi, Viktor and Pyzer-Knapp, Edward O},
  title = {Hessian QM9: A quantum chemistry database of molecular Hessians in implicit solvents},
  journal = {Scientific Data},
  year = {2025},
  volume = {12},
  number = {1},
  pages = {9},
  publisher = {Nature Publishing Group UK London},
  doi = {10.1038/s41597-024-04361-2}
}

@misc{koker2026pft,
  author = {Koker, Teddy and Gangan, Abhijeet and Kotak, Mit and Marian, Jaime and Smidt, Tess},
  title = {PFT: Phonon Fine-tuning for Machine Learned Interatomic Potentials},
  year = {2026},
  doi = {10.48550/arXiv.2601.07742},
  archive = {arXiv},
  note = {arXiv:2601.07742}
}

@inproceedings{amin2025towards,
  author = {Amin, Ishan and Raja, Sanjeev and Krishnapriyan, Aditi},
  title = {Towards Fast, Specialized Machine Learning Force Fields: Distilling Foundation Models via Energy Hessians},
  booktitle = {The Thirteenth International Conference on Learning Representations},
  year = {2025},
  doi = {10.48550/arXiv.2501.09009},
  archive = {arXiv},
  note = {ICLR 2025. arXiv:2501.09009}
}

@article{saibaba2017randomized,
  author = {Saibaba, Arvind K and Alexanderian, Alen and Ipsen, Ilse CF},
  title = {Randomized matrix-free trace and log-determinant estimators},
  journal = {Numerische Mathematik},
  year = {2017},
  volume = {137},
  number = {2},
  pages = {353--395},
  publisher = {Springer}
}

@article{herbold2022hessian,
  author = {Herbold, Marius and Behler, J{\"o}rg},
  title = {A hessian-based assessment of atomic forces for training machine learning interatomic potentials},
  journal = {The Journal of Chemical Physics},
  year = {2022},
  volume = {156},
  number = {11},
  pages = {114106},
  doi = {10.1063/5.0082952},
  publisher = {AIP Publishing}
}

@article{kohn1996nearsightedness,
  author = {Kohn, W.},
  title = {Density Functional and Density Matrix Method Scaling Linearly with the Number of Atoms},
  journal = {Physical Review Letters},
  year = {1996},
  volume = {76},
  pages = {3168--3171},
  publisher = {American Physical Society},
  doi = {10.1103/PhysRevLett.76.3168},
  month = {Apr},
  issue = {17},
  numpages = {0}
}

@book{griewank2008evaluating,
  author = {Griewank, Andreas and Walther, Andrea},
  title = {Evaluating derivatives: principles and techniques of algorithmic differentiation},
  year = {2008},
  publisher = {SIAM}
}

@article{baydin2018automatic,
  author = {Baydin, Atilim Gunes and Pearlmutter, Barak A and Radul, Alexey Andreyevich and Siskind, Jeffrey Mark},
  title = {Automatic differentiation in machine learning: a survey},
  journal = {Journal of Machine Learning Research},
  year = {2018},
  volume = {18},
  number = {153},
  pages = {1--43}
}

@article{unke2021machine,
  author = {Unke, Oliver T and Chmiela, Stefan and Sauceda, Huziel E and Gastegger, Michael and Poltavsky, Igor and Sch\"utt, Kristof T. and Tkatchenko, Alexandre and M\"uller, Klaus-Robert},
  title = {Machine learning force fields},
  journal = {Chemical Reviews},
  year = {2021},
  volume = {121},
  number = {16},
  pages = {10142--10186},
  publisher = {ACS Publications}
}

@article{behler2007generalized,
  title = {Generalized Neural-Network Representation of High-Dimensional Potential-Energy Surfaces},
  author = {Behler, J\"org and Parrinello, Michele},
  journal = {Physical Review Letters},
  volume = {98},
  pages = {146401},
  numpages = {4},
  year = {2007},
  month = {Apr},
  publisher = {American Physical Society},
  doi = {10.1103/PhysRevLett.98.146401},
}

@article{Smith2017-1,
  author = {Smith, J. S. and Isayev, O. and Roitberg, A. E.},
  title = {{ANI-1: an extensible neural network potential with DFT accuracy at force field computational cost}},
  journal = {Chemical Science},
  year = {2017},
  volume = {8},
  number = {4},
  pages = {3192--3203},
  publisher = {Royal Society of Chemistry},
  doi = {10.1039/C6SC05720A},
  archive = {arXiv}
}

@article{Smith2017-2,
  author = {Smith, J. S. and Isayev, O. and Roitberg, A. E.},
  title = {{ANI-1, A data set of 20 million calculated off-equilibrium conformations for organic molecules}},
  journal = {Scientific Data},
  year = {2017},
  volume = {4},
  number = {1},
  pages = {1--8},
  publisher = {Nature},
  doi = {10.1038/sdata.2017.193}
}

@article{Smith2018,
  author = {Smith, Justin S. and Nebgen, Ben and Lubbers, Nicholas and Isayev, Olexandr and Roitberg, Adrian E.},
  title = {Less is more: Sampling chemical space with active learning},
  journal = {The Journal of Chemical Physics},
  year = {2018},
  volume = {148},
  number = {24},
  pages = {241733},
  doi = {10.1063/1.5023802}
}

@article{Smith2020,
  author = {Smith, Justin S. and Zubatyuk, Roman and Nebgen, Benjamin and Lubbers, Nicholas and Barros, Kipton and Roitberg, Adrian E. and Isayev, Olexandr and Tretiak, Sergei},
  title = {{The ANI-1ccx and ANI-1x data sets, coupled-cluster and density functional theory properties for molecules}},
  journal = {Scientific Data},
  year = {2020},
  volume = {7},
  number = {1},
  pages = {1--10},
  doi = {10.1038/s41597-020-0473-z}
}

@article{Devereux2020,
  author = {Devereux, Christian and Smith, Justin S. and Huddleston, Kate K. and Barros, Kipton and Zubatyuk, Roman and Isayev, Olexandr and Roitberg, Adrian E.},
  title = {Extending the Applicability of the ANI Deep Learning Molecular Potential to Sulfur and Halogens},
  journal = {Journal of Chemical Theory and Computation},
  year = {2020},
  volume = {16},
  number = {7},
  pages = {4192--4202},
  doi = {10.1021/acs.jctc.0c00121}
}

@article{Gao2020,
  author = {Gao, Xiang and Ramezanghorbani, Farhad and Isayev, Olexandr and Smith, Justin S and Roitberg, Adrian E},
  title = {TorchANI: A Free and Open Source PyTorch-Based Deep Learning Implementation of the ANI Neural Network Potentials},
  journal = {Journal of Chemical Information and Modeling},
  year = {2020},
  volume = {60},
  number = {7},
  pages = {3408--3415},
  publisher = {ACS Publications},
  doi = {10.1021/acs.jcim.0c00451}
}

@article{lubbers2018hierarchical,
  author = {Lubbers, Nicholas and Smith, Justin S and Barros, Kipton},
  title = {Hierarchical modeling of molecular energies using a deep neural network},
  journal = {The Journal of Chemical Physics},
  year = {2018},
  volume = {148},
  number = {24},
  pages = {241715},
  doi = {10.1063/1.5023802},
  publisher = {AIP Publishing}
}

@article{zhang2018deep,
  author = {Zhang, Linfeng and Han, Jiequn and Wang, Han and Car, Roberto and E, Weinan},
  title = {Deep potential molecular dynamics: a scalable model with the accuracy of quantum mechanics},
  journal = {Physical Review Letters},
  year = {2018},
  volume = {120},
  number = {14},
  pages = {143001},
  publisher = {APS}
}

@article{bartok2010gaussian,
  author = {Bart{\'o}k, Albert P and Payne, Mike C and Kondor, Risi and Cs{\'a}nyi, G{\'a}bor},
  title = {Gaussian approximation potentials: The accuracy of quantum mechanics, without the electrons},
  journal = {Physical Review Letters},
  year = {2010},
  volume = {104},
  number = {13},
  pages = {136403},
  publisher = {APS}
}

@article{thompson2015spectral,
  author = {Thompson, Aidan P and Swiler, Laura P and Trott, Christian R and Foiles, Stephen M and Tucker, Garritt J},
  title = {Spectral neighbor analysis method for automated generation of quantum-accurate interatomic potentials},
  journal = {Journal of Computational Physics},
  year = {2015},
  volume = {285},
  pages = {316--330},
  publisher = {Elsevier}
}

@article{shapeev2016moment,
  author = {Shapeev, Alexander V},
  title = {Moment tensor potentials: A class of systematically improvable interatomic potentials},
  journal = {Multiscale Modeling \& Simulation},
  year = {2016},
  volume = {14},
  number = {3},
  pages = {1153--1173},
  publisher = {SIAM}
}

@article{schutt2018schnet,
  author = {Sch{\"u}tt, Kristof T and Sauceda, Huziel E and Kindermans, P-J and Tkatchenko, Alexandre and M{\"u}ller, K-R},
  title = {Schnet--a deep learning architecture for molecules and materials},
  journal = {The Journal of Chemical Physics},
  year = {2018},
  volume = {148},
  number = {24},
  pages = {241722},
  doi = {10.1063/1.5019779},
  publisher = {AIP Publishing}
}

@article{batzner2022NequIP,
  author = {Batzner, Simon and Musaelian, Albert and Sun, Lixin and Geiger, Mario and Mailoa, Jonathan P and Kornbluth, Mordechai and Molinari, Nicola and Smidt, Tess E and Kozinsky, Boris},
  title = {E (3)-equivariant graph neural networks for data-efficient and accurate interatomic potentials},
  journal = {Nature Communications},
  year = {2022},
  volume = {13},
  number = {1},
  pages = {2453},
  publisher = {Nature Publishing Group UK London}
}

@inproceedings{schutt2021equivariant,
  author = {Sch\"utt, Kristof and Unke, Oliver and Gastegger, Michael},
  title = {Equivariant message passing for the prediction of tensorial properties and molecular spectra},
  booktitle = {International Conference on Machine Learning},
  year = {2021},
  pages = {9377--9388},
}

@article{batatia2022mace,
  author = {Batatia, Ilyes and Kovacs, David P and Simm, Gregor and Ortner, Christoph and Cs{\'a}nyi, G{\'a}bor},
  title = {MACE: Higher order equivariant message passing neural networks for fast and accurate force fields},
  journal = {Advances in Neural Information Processing Systems},
  year = {2022},
  volume = {35},
  pages = {11423--11436}
}

@article{chmiela2019sgdml,
  author = {Chmiela, Stefan and Sauceda, Huziel E and Poltavsky, Igor and M{\"u}ller, Klaus-Robert and Tkatchenko, Alexandre},
  title = {sGDML: Constructing accurate and data efficient molecular force fields using machine learning},
  journal = {Computer Physics Communications},
  year = {2019},
  volume = {240},
  pages = {38--45},
  publisher = {Elsevier}
}

@article{unke2019physnet,
  author = {Unke, Oliver T and Meuwly, Markus},
  title = {PhysNet: A neural network for predicting energies, forces, dipole moments, and partial charges},
  journal = {Journal of Chemical Theory and Computation},
  year = {2019},
  volume = {15},
  number = {6},
  pages = {3678--3693},
  publisher = {ACS Publications}
}

@misc{burger2025shoot,
  author = {Burger, Andreas and Thiede, Luca and R{\o}nne, Nikolaj and Bernales, Varinia and Vijaykumar, Nandita and Vegge, Tejs and Bhowmik, Arghya and Aspuru-Guzik, Alan},
  title = {Shoot from the HIP: Hessian Interatomic Potentials without derivatives},
  year = {2025},
  doi = {10.48550/arXiv.2509.21624},
  archive = {arXiv},
  note = {arXiv:2509.21624}
}

@inproceedings{hardt2016train,
  author    = {Hardt, Moritz and Recht, Benjamin and Singer, Yoram},
  title     = {Train faster, generalize better: Stability of stochastic gradient descent},
  booktitle = {Proceedings of the 33rd International Conference on Machine Learning},
  volume    = {48},
  pages     = {1225--1234},
  year      = {2016}
}

@article{hochreiter1997flat,
  author = {Hochreiter, Sepp and Schmidhuber, J{\"u}rgen},
  title = {Flat minima},
  journal = {Neural computation},
  year = {1997},
  volume = {9},
  number = {1},
  pages = {1--42},
  publisher = {MIT Press One Rogers Street, Cambridge, MA 02142-1209, USA journals-info~…}
}

@article{pulay2014analytical,
  author = {Pulay, Peter},
  title = {Analytical derivatives, forces, force constants, molecular geometries, and related response properties in electronic structure theory},
  journal = {Wiley Interdisciplinary Reviews: Computational Molecular Science},
  year = {2014},
  volume = {4},
  number = {3},
  pages = {169--181},
  publisher = {Wiley Online Library}
}

@article{pople1979derivative,
  author = {Pople, J\_A and Krishnan, R and Schlegel, HB and Binkley, J S\_},
  title = {Derivative studies in hartree-fock and m{\o}ller-plesset theories},
  journal = {International Journal of Quantum Chemistry},
  year = {1979},
  volume = {16},
  number = {S13},
  pages = {225--241},
  publisher = {Wiley Online Library}
}

@article{gonzalez1989improved,
  author = {Gonzalez, Carlos and Schlegel, H Bernhard},
  title = {An improved algorithm for reaction path following},
  journal = {The Journal of Chemical Physics},
  year = {1989},
  volume = {90},
  number = {4},
  pages = {2154--2161},
  publisher = {American Institute of Physics}
}

@article{fukui1981path,
  author = {Fukui, Kenichi},
  title = {The path of chemical reactions-the IRC approach},
  journal = {Accounts of Chemical Research},
  year = {1981},
  volume = {14},
  number = {12},
  pages = {363--368},
  publisher = {ACS Publications}
}

@inproceedings{martens2010deep,
  author = {Martens, James and others},
  title = {Deep learning via hessian-free optimization.},
  booktitle = {Proceedings of the 27th International Conference on Machine Learning},
  year = {2010},
  volume = {27},
  pages = {735--742}
}

@book{nocedal2006numerical,
  author = {Nocedal, Jorge and Wright, Stephen J},
  title = {Numerical optimization},
  year = {2006},
  publisher = {Springer}
}

@article{halko2011finding,
  author = {Halko, N. and Martinsson, P. G. and Tropp, J. A.},
  title = {Finding Structure with Randomness: Probabilistic Algorithms for Constructing Approximate Matrix Decompositions},
  journal = {SIAM Review},
  year = {2011},
  volume = {53},
  number = {2},
  pages = {217-288},
  doi = {10.1137/090771806}
}

@misc{woodruff2014sketching,
  author  = {Woodruff, David P.},
  title   = {Sketching as a tool for numerical linear algebra},
  year    = {2014},
  archive = {arXiv},
  note    = {arXiv:1411.4357}
}

@article{hutchinson1990stochastic,
  author = {M.F. Hutchinson},
  title = {A stochastic estimator of the trace of the influence matrix for laplacian smoothing splines},
  journal = {Communications in Statistics - Simulation and Computation},
  year = {1990},
  volume = {19},
  number = {2},
  pages = {433--450},
  publisher = {Taylor \& Francis},
  doi = {10.1080/03610919008812866}
}

@article{pearlmutter1994fast,
  author = {Pearlmutter, Barak A},
  title = {Fast exact multiplication by the Hessian},
  journal = {Neural computation},
  year = {1994},
  volume = {6},
  number = {1},
  pages = {147--160},
  publisher = {MIT Press}
}

@article{loew2025universal,
  author = {Loew, Antoine and Sun, Dewen and Wang, Hai-Chen and Botti, Silvana and Marques, Miguel AL},
  title = {Universal machine learning interatomic potentials are ready for phonons},
  journal = {npj Computational Materials},
  year = {2025},
  volume = {11},
  number = {1},
  pages = {178},
  publisher = {Nature Publishing Group UK London}
}

@article{avron2011randomized,
  author = {Avron, Haim and Toledo, Sivan},
  title = {Randomized algorithms for estimating the trace of an implicit symmetric positive semi-definite matrix},
  year = {2011},
  issue_date = {April 2011},
  publisher = {Association for Computing Machinery},
  address = {New York, NY, USA},
  volume = {58},
  number = {8},
  doi = {10.1145/1944345.1944349},
  journal = {Journal of the ACM},
  pages = {8:1--8:17}
}

@misc{Gaussian16,
    author={M. J. Frisch and G. W. Trucks and H. B. Schlegel and G. E. Scuseria and M. A. Robb and J. R. Cheeseman and G. Scalmani and V. Barone and G. A. Petersson and H. Nakatsuji and X. Li and M. Caricato and A. V. Marenich and J. Bloino and B. G. Janesko and R. Gomperts and B. Mennucci and H. P. Hratchian and J. V. Ortiz and A. F. Izmaylov and J. L. Sonnenberg and D. Williams-Young and F. Ding and F. Lipparini and F. Egidi and J. Goings and B. Peng and A. Petrone and T. Henderson and D. Ranasinghe and V. G. Zakrzewski and J. Gao and N. Rega and G. Zheng and W. Liang and M. Hada and M. Ehara and K. Toyota and R. Fukuda and J. Hasegawa and M. Ishida and T. Nakajima and Y. Honda and O. Kitao and H. Nakai and T. Vreven and K. Throssell and Montgomery, {Jr.}, J. A. and J. E. Peralta and F. Ogliaro and M. J. Bearpark and J. J. Heyd and E. N. Brothers and K. N. Kudin and V. N. Staroverov and T. A. Keith and R. Kobayashi and J. Normand and K. Raghavachari and A. P. Rendell and J. C. Burant and S. S. Iyengar and J. Tomasi and M. Cossi and J. M. Millam and M. Klene and C. Adamo and R. Cammi and J. W. Ochterski and R. L. Martin and K. Morokuma and O. Farkas and J. B. Foresman and D. J. Fox},
    title={Gaussian˜16 {R}evision {C}.01},
    year={2016},
    note={Gaussian Inc. Wallingford CT}
}

\newpage
\appendix
\section{Projected Hessian Learning Training Algorithm}
\label{app:phl_algorithm}

\begin{algorithm}[h!]
\DontPrintSemicolon
\caption{Projected Hessian Learning (PHL) for second-order derivative training of MLIPs using Hessian-Vector Products (HVPs)}
\label{alg:phl_training}
\KwIn{Dataset $\mathcal{D}=\{(\mathbf{R},E,\mathbf{F},\mathcal{C})\}$, where $\mathbf{R}\in\mathbb{R}^{3N}$, $E$ is reference energy, $\mathbf{F}=-\nabla_{\mathbf{R}}E$ is reference force. Curvature data $\mathcal{C}$ may contain one or more HVP pairs $(v,\,Hv)$.}
\KwIn{MLIP $\tilde{E}_\theta(\mathbf{R})$, loss weights $\lambda_E,\lambda_F,\lambda_H$, probes per structure $K$, step size $\eta$.}
\KwOut{Trained parameters $\theta$.}

\BlankLine
Initialize parameters $\theta$\;
\While{not converged}{
  Sample minibatch $\mathcal{B}\subset\mathcal{D}$\;
  $\mathcal{L}\leftarrow 0$\;
  \ForEach{$(\mathbf{R},E,\mathbf{F},\mathcal{C})\in\mathcal{B}$}{
    $\tilde{E}\leftarrow \tilde{E}_\theta(\mathbf{R})$\;
    $\tilde{\mathbf{F}}\leftarrow -\nabla_{\mathbf{R}}\tilde{E}$\;
    $\mathcal{L}_E \leftarrow \|\tilde{E}-E\|^2$\;
    $\mathcal{L}_F \leftarrow \frac{1}{3N}\|\tilde{\mathbf{F}}-\mathbf{F}\|^2$\;

    \tcp*[l]{PHL curvature loss: average over $K$ Hessian--vector probes}
    $\hat{\mathcal{L}}_H \leftarrow 0$\;
    \For{$k\leftarrow 1$ \KwTo $K$}{
      \uIf{$(v_k,\,Hv_k)\in\mathcal{C}$}{
        $v\leftarrow v_k$\;
        $y\leftarrow Hv_k$\;
      }
      \Else{
        \tcp*[l]{Generate Gaussian probe and obtain reference $y=Hv$}
        Sample $v\sim\mathcal{N}(0,I)$\;
        Obtain reference $y\leftarrow Hv$ using the available QC procedure\;
      }
      \tcp*[l]{Compute predicted HVP without forming $\tilde{H}$}
      $\tilde{y}\leftarrow \tilde{H}v$ via AD HVP (e.g.\ $\tilde{y}=\nabla_{\mathbf{R}}(\tilde{\mathbf{F}})\,v$)\;
      $\hat{\mathcal{L}}_{H} \leftarrow \hat{\mathcal{L}}_{H} + \frac{1}{(3N)^2}\|\tilde{y}-y\|^2$\;
    }
    $\hat{\mathcal{L}}_{H} \leftarrow \frac{1}{K}\hat{\mathcal{L}}_{H}$\;

    $\mathcal{L}\leftarrow \mathcal{L} + \lambda_E\mathcal{L}_E + \lambda_F\mathcal{L}_F + \lambda_H\hat{\mathcal{L}}_{H}$\;
  }
  $\theta \leftarrow \theta - \eta\,\nabla_{\theta}\mathcal{L}$\;
}
\end{algorithm}


\clearpage  

\renewcommand{\thesection}{\Alph{section}}  
\renewcommand{\thefigure}{S\arabic{figure}}  
\renewcommand{\thetable}{S\arabic{table}}  
\renewcommand{\theequation}{S\arabic{equation}}  

\setcounter{section}{0}   
\setcounter{figure}{0}    
\setcounter{table}{0}     
\setcounter{equation}{0}  

\begin{center}
    {\LARGE\textbf{Supplementary Information\\[0.5em]
    \Large Projected Hessian Learning: Fast Curvature Supervision for Accurate Machine-Learning Interatomic Potentials}}
\end{center}

\section{Stochastic error analysis}

Although $\left\langle v_{i}v_{j}\right\rangle =\delta_{ij}$ guarantees an unbiased approximation, the accuracy of the approximation can vary significantly according to the distribution of $v$. We follow Avron \emph{et al.},\cite{avron2011randomized} and analyze the mean squared error of the stochastic estimator $\mathrm{tr}\,A\approx v^{T}Av$,

\begin{align}
\textrm{MSE}[v] & =\left\langle \left(\mathrm{tr}\,A-v^{T}Av\right)^{2}\right\rangle .\label{eq:MSE}
\end{align}
Since we are working with unbiased estimators,

\begin{align}
\mathrm{tr}\,A=\sum_{i}A_{ii}=\sum_{ij}\delta_{ij}A_{ij}=\sum_{ij}\left\langle v_{i}v_{j}\right\rangle A_{ij}=\left\langle v^{T}Av\right\rangle .\label{eq:si_unbiased}
\end{align}
this becomes

\begin{align}
\textrm{MSE}[v] & =\left(\mathrm{tr}\,A\right)^{2}-2\mathrm{tr}\,A\left\langle v^{T}Av\right\rangle +\left\langle \left(v^{T}Av\right)^{2}\right\rangle \nonumber \\
 & =\left\langle \left(v^{T}Av\right)^{2}\right\rangle -\left(\mathrm{tr}\,A\right)^{2}.
 \label{eq:err_expand}
\end{align}
It remains to evaluate $\left\langle\left(v^{T}Av\right)^{2}\right\rangle$:

\begin{align}
\left\langle \left(v^{T}Av\right)^{2}\right\rangle  & =\left\langle \sum_{ij}v_{i}A_{ij}v_{j}\sum_{kl}v_{k}A_{kl}v_{l}\right\rangle \nonumber \\
 & =\sum_{ijkl}A_{ij}A_{kl}\left\langle v_{i}v_{j}v_{k}v_{l}\right\rangle .
\end{align}
Observe that the random vector $v$ enters the error entirely through the four-point expectation $\left\langle v_{i}v_{j}v_{k}v_{l}\right\rangle $.

Let us first consider the Hutchinson estimator $v_{i}^{\textrm{Hutch}}=\{\pm1\}\label{eq:si_hutch}$. There are only two possible values,
\[
\left\langle v_{i}^{\textrm{Hutch}}v_{j}^{\textrm{Hutch}}v_{k}^{\textrm{Hutch}}v_{l}^{\textrm{Hutch}}\right\rangle =\begin{cases}
1 & \textrm{indices in even groups}\\
0 & \textrm{otherwise}
\end{cases}.
\]
The first branch includes four possible cases: (1) the indices are all equal, $i=j=k=l$, or the indices come in two equal pairs, (2) $i=j\neq k=l$, (3) $i=k\neq j=l$, or (4) $i=l\neq j=k$. Using the fact that $A$ is symmetric, cases (3) and (4) are equivalent and can be merged. This yields three terms;

\begin{equation}
\left\langle \left(v^{\textrm{Hutch},T}Av^{\textrm{Hutch}}\right)^{2}\right\rangle =\sum_{i}A_{ii}^{2}+\sum_{i\neq k}A_{ii}A_{kk}+2\sum_{i\neq j}A_{ij}^{2}.
\end{equation}
One can also expand this definition.

\begin{equation}
\left(\mathrm{tr}\,A\right)^{2}=\left(\sum_{i}A_{ii}\right)^{2}=\sum_{i}A_{ii}^{2}+\sum_{i\neq k}A_{ii}A_{kk}.
\label{eq:trA2_expand}
\end{equation}
Subtraction yields the mean squared error for the Hutchinson estimator,

\begin{equation}
\textrm{MSE}[v^{\mathrm{Hutch}}]=2\sum_{i\neq j}A_{ij}^{2}.
\label{eq:hutch_err}
\end{equation}
Now, let us turn to the 1-hot encoding distribution,

\begin{equation}
v_{i}^{\textrm{1Hot}}=\sqrt{3N}\,\delta_{i,c}\label{eq:si_1hot}
\end{equation}
In this case,

\begin{equation}
v_{i}^{\textrm{1Hot}}v_{j}^{\textrm{1Hot}}v_{k}^{\textrm{1Hot}}v_{l}^{\textrm{1Hot}}=\begin{cases}
(3N)^{2} & \textrm{all indices are the hot index}\\
0 & \textrm{otherwise}
\end{cases}.
\end{equation}
Recall that, in the context of Hessian training, $N$ is the number of atoms and $3N$ is our matrix dimension. There is a $1/3N$ chance that a given index is the randomized hot index, so

\begin{equation}
\left\langle v_{i}^{\textrm{1Hot}}v_{j}^{\textrm{1Hot}}v_{k}^{\textrm{1Hot}}v_{l}^{\textrm{1Hot}}\right\rangle =\begin{cases}
3N & \textrm{all indices equal}\\
0 & \textrm{otherwise}
\end{cases}.
\end{equation}
It follows that

\begin{equation}
\left\langle \left(v^{\textrm{1Hot},T}Av^{\textrm{1Hot}}\right)^{2}\right\rangle =3N\sum_{i}A_{ii}^{2}
\end{equation}
Using Eqs. (\ref{eq:err_expand}) and (\ref{eq:trA2_expand}) we arrive at the mean-squared error of the 1-hot column estimator

\begin{equation}
\textrm{MSE}[v^{\mathrm{1Hot}}]=(3N-1)\sum_{i}A_{ii}^{2}-\sum_{i, k ; i\neq k}A_{ii}A_{kk}.
\label{eq:1hot_err}
\end{equation}

In our setting, $A=B^{T}B$, where $B$ represents the error in the Hessian predicted by our Machine Learning (ML) model,
\begin{align}
A & =B^{T}B\\
B & =\frac{\tilde{H}-H}{3N}.\label{eq:si_B}
\end{align}
Physical Hessians are expected to decay rapidly with interatomic distance, $|\mathbf{r}_i-\mathbf{r}_j|$, a behavior observed empirically and consistent with the locality assumptions underlying most MLIP architectures.\cite{kohn1996nearsightedness,herbold2022hessian} Consequently, only $O(N)$ matrix elements of $A$ are expected to contribute appreciably in the limit of the system size $N$.

Under this locality assumption, the Hutchinson estimator yields a mean-squared error (Eq.~(\ref{eq:hutch_err})) that scales as $O(N)$ for extensive quantities, implying an RMSE that grows as $\sqrt{N}$ and a relative error that decreases with increasing system size. By contrast, for one-hot random vectors (Eq.~(\ref{eq:si_1hot})), all contributions to the mean-squared error (Eq.~(\ref{eq:1hot_err})) arise from diagonal elements $A_{ii}$. Although these terms are likewise localized, the resulting MSE scales as $O(N^{2})$, leading to less favorable asymptotic behavior for large systems compared to Hutchinson under the same locality assumptions.

\section{Training Procedure}

The models were trained using mini-batch gradient descent with adaptive learning-rate schedules. The maximum number of epochs was set to 5000, although this limit was never reached in practice. Each method was trained as an ensemble of five independent models with different random seeds, and the reported RMSE values correspond to the ensemble means with error bars representing standard deviations.

\begin{itemize}
    \item \textbf{Dataset split:} For the reactant, transition state, and product dataset, 80\% of the structures were used for training, 10\% for validation and 10\% were held as a benchmark test set.
    \item \textbf{Optimizers:} Model weights were updated using the AdamW optimizer, while biases were updated with stochastic gradient descent (SGD). Both optimizers were set to a learning rate of \SI{1e-4}, $\beta_1=0.9$, $\beta_2=0.999$, $eps=$ \SI{1e-8}, and a weight decay of \SI{1e-2}.
    \item \textbf{Batch composition:} Each minibatch contained 400 molecular structures drawn from a dataset of reactants, transition states, and products, ensuring coverage of equilibrium configurations.
    \item \textbf{Hutchinson estimator (PHL):} Probe vectors were sampled as independent Gaussian random variables $v_{i}^{\textrm{Hutch}}$, which satisfy $\langle v_{i} \rangle = 0$ and $\langle v_{i}v_{j} \rangle = \delta_{ij}$. This ensures that the estimator is unbiased, i.e., \ $\mathbb{E}[v^{T}Av] = \mathrm{tr}(A)$. 
    \item \textbf{One-column estimator:} Following recent work in Hessian training,\cite{cui2025large} probe vectors were chosen to be one-hot basis vectors $v_{i}^{\textrm{1Hot}}$ scaled by $\sqrt{3N}$. Only one element in each probe vector is set to one, while the rest are set to zero. This approach corresponds to selecting a single column of the Hessian at random. In mathematical terms, 
    $$v_{i}^{\textrm{1Hot}} = \sqrt{3N}\,\delta_{i,c} \qquad c \in \{1,\dots,3N\}$$
    \item \textbf{Framework:} The models were implemented in PyTorch with custom operators for force and Hessian backpropagation.
    \item \textbf{Hardware:} Training was performed on various models of NVIDIA GPUs in high performance computing clusters.
    \item \textbf{Training runtime:} Wall-clock time per epoch was recorded for each training method (E–F, E–F–HVP One-Column, E–F–HVP PHL, and E–F–H) on NVIDIA A6000 GPUs. The reported values correspond to averages over multiple epochs.  
    \item \textbf{DFT data generation:} CPU scaling of reference calculations was evaluated using Gaussian16 for molecular systems containing up to 100 atoms.
\end{itemize}

\section{Bland--Altman Analysis of PHL Probing Strategies}
\label{sec:si_bland_altman}

This section compares two methods that differ only in the distribution used to probe curvature through Hessian-vector products (HVPs): (i) one-column (one-hot) probing, which samples a single Hessian column per probe, and (ii) PHL probing (Hutchinson-type randomized projections), which aggregates information across many curvature directions in expectation. For each metric (energy, force, and Hessian RMSE), we use Bland-Altman plots to visualize the paired differences between methods, defined throughout as
\[
\Delta \mathrm{RMSE} \equiv \mathrm{RMSE}_{\text{one-column}}-\mathrm{RMSE}_{\text{PHL}}
\]
Thus, a positive $\Delta\mathrm{RMSE}$ indicates that Hutchinson-based PHL is more accurate.

\subsection{Randomized HVP Probes per Minibatch}
\label{subsec:si_randomized_probes}

\begin{figure}[H]
    \centering
    {\includegraphics[scale=0.50]{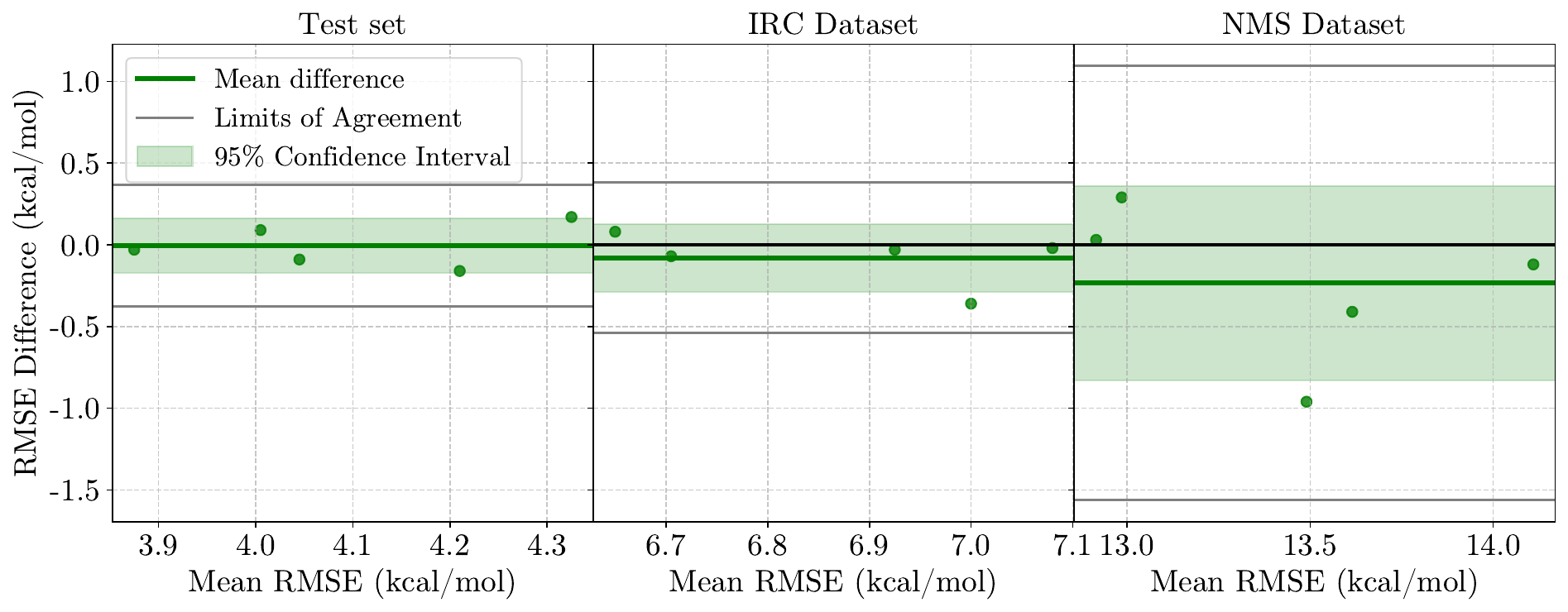}}
    \caption{Bland--Altman analysis of energy RMSE differences for randomized probes per minibatch. Differences are defined as $\Delta \mathrm{RMSE}=\mathrm{RMSE}_{\text{one-column}}-\mathrm{RMSE}_{\text{PHL}}$ across the Test, IRC, and NMS datasets. Each point represents a paired set of trained models. The green line denotes the mean difference and the shaded region its 95\% CI; black lines indicate limits of agreement. The 95\% CI includes zero for all datasets, indicating no statistically significant difference in energy accuracy between probing strategies when probes are randomized each minibatch.}
    \label{fig:si_bland_altman_energy_randomized_v}
\end{figure}

\begin{figure}[H]
    \centering
    {\includegraphics[scale=0.50]{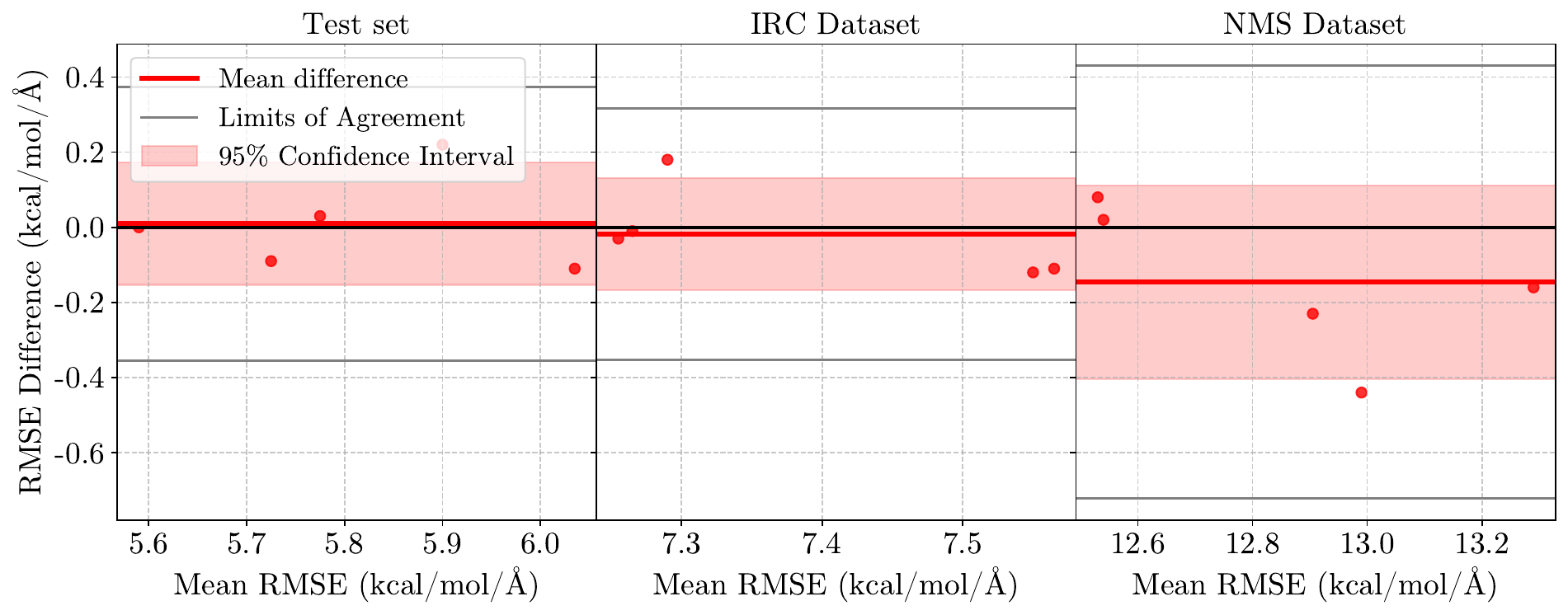}}
    \caption{Bland--Altman analysis of force RMSE differences for randomized probes per minibatch, with $\Delta \mathrm{RMSE}$ defined as in Fig.~\ref{fig:si_bland_altman_energy_randomized_v}. The 95\% CI includes zero for all datasets, indicating no statistically significant difference in force accuracy between one-column and PHL methods when probes are randomized each minibatch.}
    \label{fig:si_bland_altman_force_randomized_v}
\end{figure}

\begin{figure}[H]
    \centering
    {\includegraphics[scale=0.50]{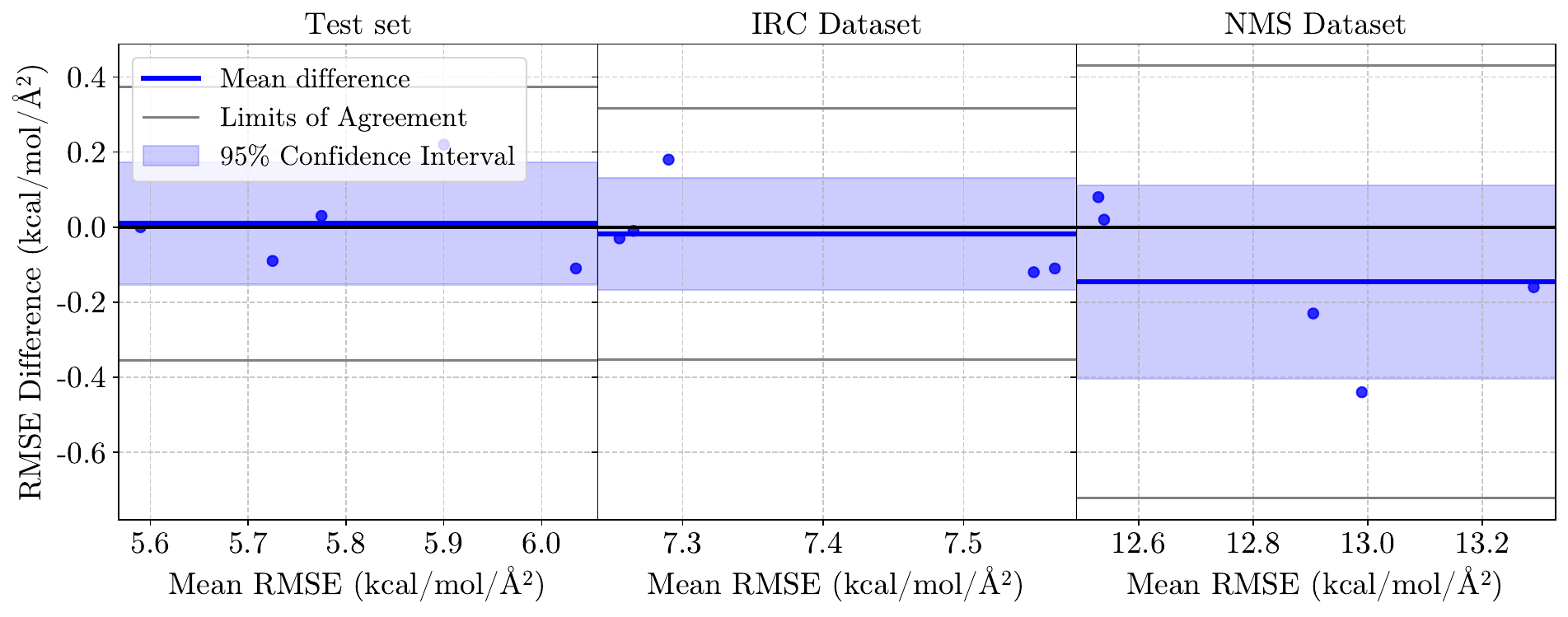}}
    \caption{Bland--Altman analysis of Hessian RMSE differences for randomized probes per minibatch, with $\Delta \mathrm{RMSE}$ defined as in Fig.~\ref{fig:si_bland_altman_energy_randomized_v}. The 95\% CI includes zero for all datasets, indicating no statistically significant difference in Hessian accuracy between probing strategies in the randomized-probe regime.}
    \label{fig:si_bland_altman_hessian_randomized_v}
\end{figure}

In the randomized-probe setting, we resample the probing vector(s) at each minibatch, so that both estimators access many independent curvature directions over the course of training. Figures~\ref{fig:si_bland_altman_energy_randomized_v}--\ref{fig:si_bland_altman_hessian_randomized_v} show that the zero line lies within the 95\% confidence interval (CI) of the mean difference across the Test, IRC, and NMS datasets for \emph{all} three metrics. Therefore, when probes are randomized per minibatch, the one-hot and Hutchinson estimators yield statistically indistinguishable accuracy, consistent with our main-text conclusion that both stochastic estimators effectively approximate full Hessian supervision in this regime for our characteristic system sizes.

\subsection{Fixed HVP Probe per System (Data-Limited Regime)}
\label{subsec:si_fixed_probes}

In the fixed-probe setting, we assign each molecular system a single probe vector (i.e., one reference HVP) that remains fixed throughout training. This mimics a data-limited regime where only restricted second-derivative information is available per structure. In this setting, Gaussian-probed PHL becomes consistently more accurate than one-column probing, with the strongest gains appearing on the extrapolative NMS dataset.

\begin{figure}[H]
    \centering
    {\includegraphics[scale=0.50]{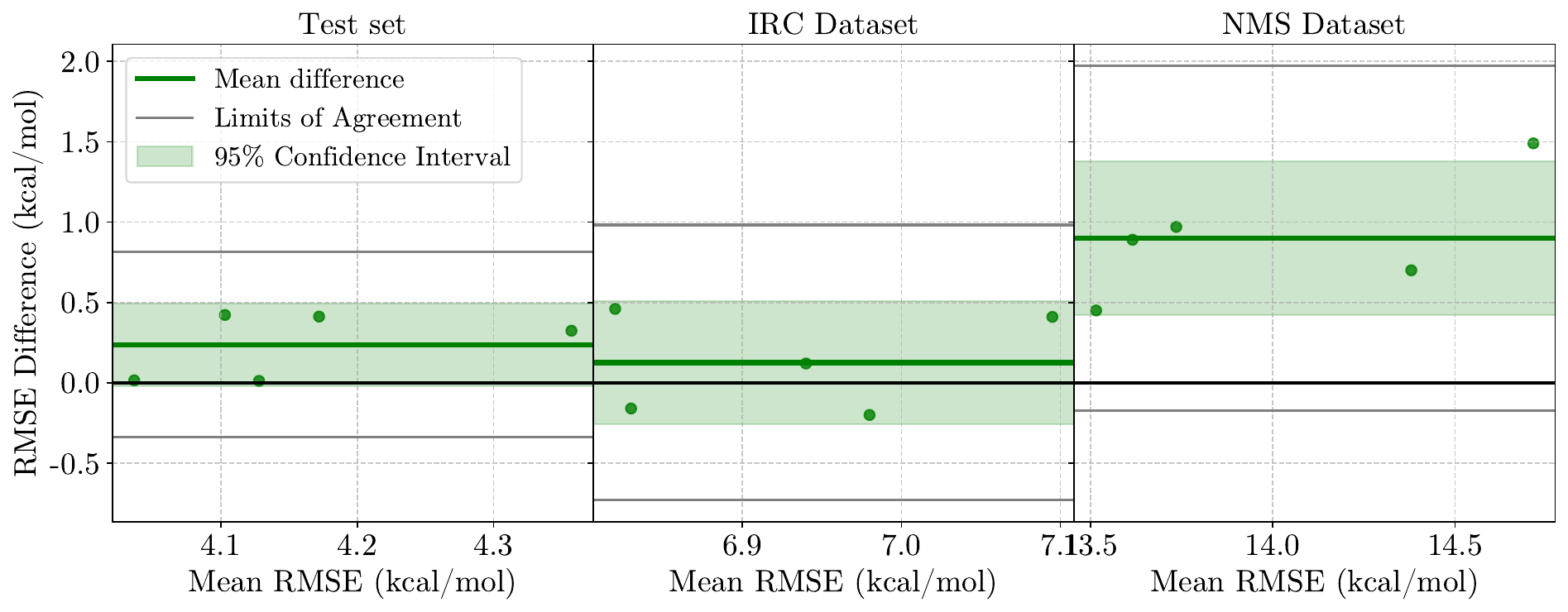}}
    \caption{Bland--Altman analysis of energy RMSE differences for fixed probes per system. Differences are defined as $\Delta \mathrm{RMSE}=\mathrm{RMSE}_{\text{one-column}}-\mathrm{RMSE}_{\text{PHL}}$ across the Test, IRC, and NMS datasets. The mean difference is not significant for Test and IRC (95\% CI includes zero), but is significantly positive for NMS (95\% CI excludes zero), indicating better energy extrapolation for Hutchinson-based PHL in the data-limited regime.}
    \label{fig:si_bland_altman_energy_fixed_v}
\end{figure}

\begin{figure}[H]
    \centering
    {\includegraphics[scale=0.50]{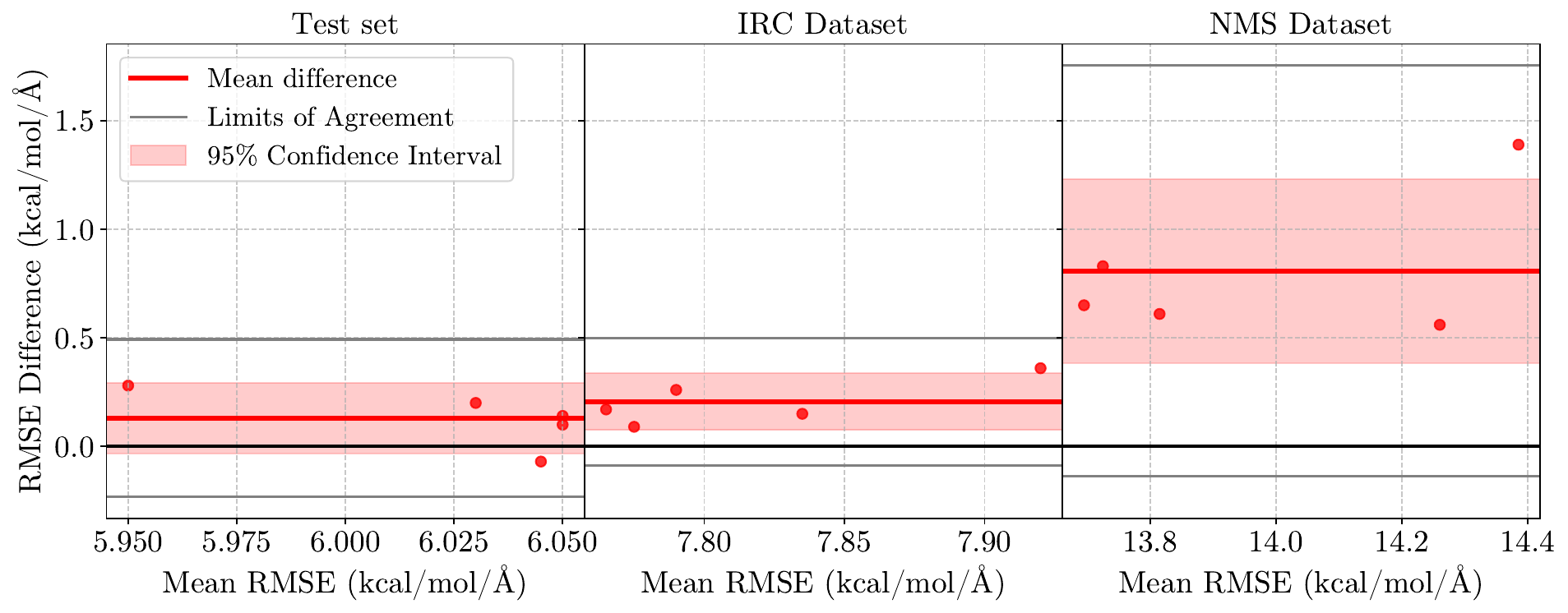}}
    \caption{Bland--Altman analysis of force RMSE differences for fixed probes per system, with $\Delta \mathrm{RMSE}$ defined as in Fig.~\ref{fig:si_bland_altman_energy_fixed_v}. The 95\% CI excludes zero for IRC and NMS, demonstrating that the one-column estimator yields significantly higher force errors than the Hutchinson-based PHL estimator on intermediate (IRC) and extrapolative (NMS) geometries.}
    \label{fig:si_bland_altman_force_fixed_v}
\end{figure}

\begin{figure}[H]
    \centering
    {\includegraphics[scale=0.50]{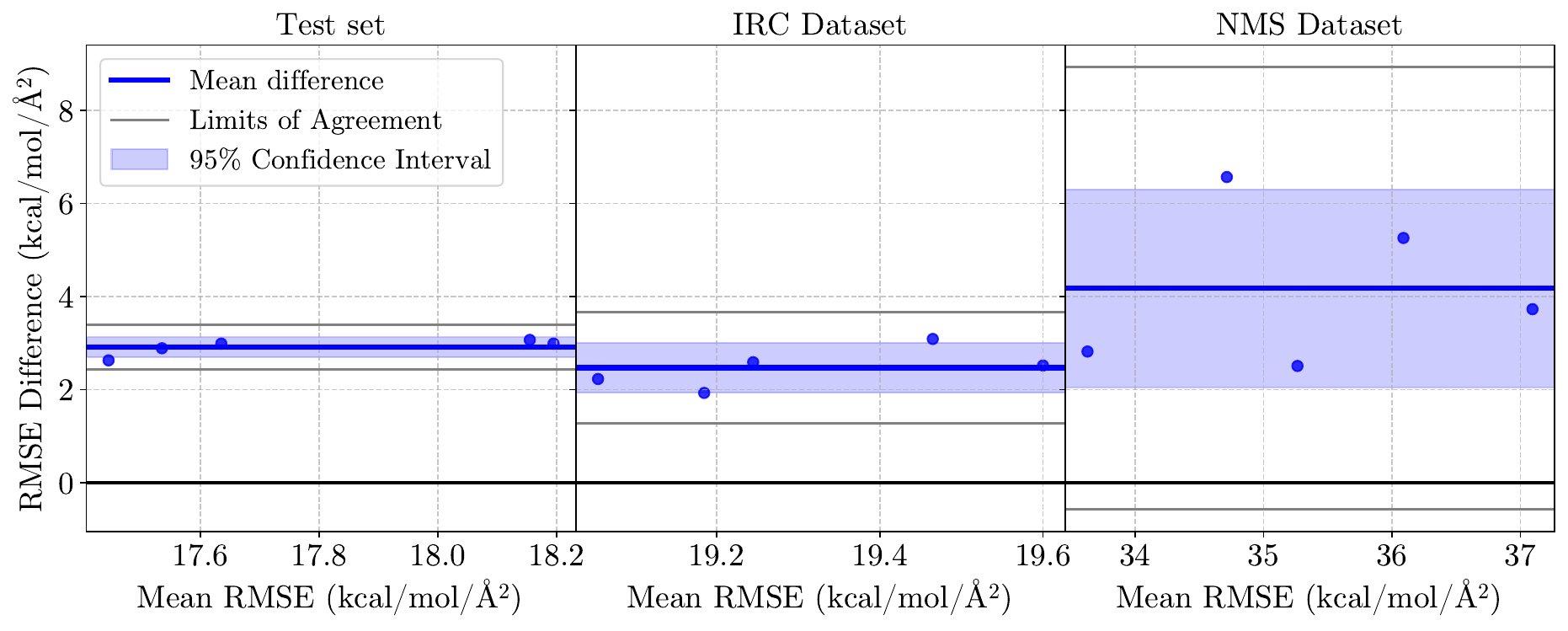}}
    \caption{Bland--Altman analysis of Hessian RMSE differences for fixed probes per system, with $\Delta \mathrm{RMSE}$ defined as in Fig.~\ref{fig:si_bland_altman_energy_fixed_v}. The 95\% CI excludes zero for Test, IRC, and NMS, indicating that the one-column estimator produces significantly higher Hessian errors than the Hutchinson-based PHL estimator across all datasets in the fixed-probe regime, with the largest discrepancies observed on NMS.}
    \label{fig:si_bland_altman_hessian_fixed_v}
\end{figure}

Figures~\ref{fig:si_bland_altman_energy_fixed_v}--\ref{fig:si_bland_altman_hessian_fixed_v} show that the mean difference shifts positively (favoring PHL), and the zero line falls outside the 95\% CI for increasingly many datasets as the target becomes more challenging. Specifically, energy differences are significant on NMS, force differences are significant on IRC and NMS, and Hessian differences are significant on all three datasets. These trends agree with our main-text statistics: under fixed-vector conditions on NMS, Hutchinson-based PHL yields additional reductions of 6.2\% (energy RMSE), 5.6\% (force RMSE) and 11.2\% (Hessian RMSE) relative to one-column probing (paired $t$-tests).

\noindent\textbf{Summary.} When HVP probes are randomized at each minibatch, one-column and PHL are statistically indistinguishable for our characteristic system size (a median of $N\sim14$). When only a single probe per system is available (data-limited regime), PHL is consistently more accurate, particularly for extrapolative geometries, establishing it as the more reliable default probing strategy for practical second-order training with limited curvature data.

\end{document}